\documentclass[12pt]{iopart}
\usepackage{graphicx}
\usepackage{textcomp}
\usepackage{dcolumn}
\usepackage{amssymb}
\usepackage{xcolor}
\usepackage[english]{babel}
\usepackage{babel}
\usepackage{nicefrac}

\begin{document}

\title[]{Long-lasting XUV activation of helium nanodroplets for avalanche ionization}

\author{C. Medina}
\address{Institute of Physics, University of Freiburg, 79104 Freiburg, Germany}

\author{A. Ø. Lægdsmand} 
\address{Department of Physics and Astronomy, Aarhus University, 8000 Aarhus C, Denmark}

\author{L. Ben Ltaief} 
\address{Department of Physics and Astronomy, Aarhus University, 8000 Aarhus C, Denmark}

\author{Z. Hoque}
\address{ELI Beamlines, The Extreme Light Infrastructure ERIC, Za Radnic\'{i} 835, 252 41, Doln\'{i} B\v{r}e\v{z}any, Czech Republic}

\author{A. H. Roos}
\address{ELI Beamlines, The Extreme Light Infrastructure ERIC, Za Radnic\'{i} 835, 252 41, Doln\'{i} B\v{r}e\v{z}any, Czech Republic}

\author{M. Jurkovi\v{c}}
\address{ELI Beamlines, The Extreme Light Infrastructure ERIC, Za Radnic\'{i} 835, 252 41, Doln\'{i} B\v{r}e\v{z}any, Czech Republic}
\address{Czech Technical University in Prague, FNSPE, B\v{r}ehov\'{a} 7, 115 19 Prague 1, Czechia}

\author{O. Hort}
\address{ELI Beamlines, The Extreme Light Infrastructure ERIC, Za Radnic\'{i} 835, 252 41, Doln\'{i} B\v{r}e\v{z}any, Czech Republic}

\author{O. Finke}
\address{ELI Beamlines, The Extreme Light Infrastructure ERIC, Za Radnic\'{i} 835, 252 41, Doln\'{i} B\v{r}e\v{z}any, Czech Republic}
\address{Czech Technical University in Prague, FNSPE, B\v{r}ehov\'{a} 7, 115 19 Prague 1, Czechia}

\author{M. Albrecht}
\address{ELI Beamlines, The Extreme Light Infrastructure ERIC, Za Radnic\'{i} 835, 252 41, Doln\'{i} B\v{r}e\v{z}any, Czech Republic}
\address{Czech Technical University in Prague, FNSPE, B\v{r}ehov\'{a} 7, 115 19 Prague 1, Czechia}

\author{J. Nejdl}
\address{ELI Beamlines, The Extreme Light Infrastructure ERIC, Za Radnic\'{i} 835, 252 41, Doln\'{i} B\v{r}e\v{z}any, Czech Republic}
\address{Czech Technical University in Prague, FNSPE, B\v{r}ehov\'{a} 7, 115 19 Prague 1, Czechia}

\author{F. Stienkemeier}
\address{Institute of Physics, University of Freiburg, 79104 Freiburg, Germany}

\author{J. Andreasson}
\address{ELI Beamlines, The Extreme Light Infrastructure ERIC, Za Radnic\'{i} 835, 252 41, Doln\'{i} B\v{r}e\v{z}any, Czech Republic}

\author{E. Klime\v{s}ov\'{a}}
\address{ELI Beamlines, The Extreme Light Infrastructure ERIC, Za Radnic\'{i} 835, 252 41, Doln\'{i} B\v{r}e\v{z}any, Czech Republic}

\author{M. Krikunova}
\address{ELI Beamlines, The Extreme Light Infrastructure ERIC, Za Radnic\'{i} 835, 252 41, Doln\'{i} B\v{r}e\v{z}any, Czech Republic}
\address{Technical University of Applied Sciences, Hochschulring 1, 15745 Wildau, Germany}

\author{A. Heidenreich}%
\address{Kimika Fakultatea, Euskal Herriko Unibertsitatea (UPV/EHU) and Donostia International Physics Center (DIPC), P.K. 1072, 20080 Donostia, Spain}
\address{IKERBASQUE, Basque Foundation for Science, 48011 Bilbao, Spain}

\author{M. Mudrich} 
\address{Department of Physics and Astronomy, Aarhus University, 8000 Aarhus C, Denmark}
\ead{mudrich@phys.au.dk}

\date{today}

\vspace{10pt}
\begin{indented}
\item[]November 2022
\end{indented}

\begin{abstract}
We study the dynamics of avalanche ionization of pure helium nanodroplets activated by a weak extreme-ultraviolet (XUV) pulse and driven by an intense near-infrared (NIR) pulse. In addition to a transient enhancement of ignition of a nanoplasma at short delay times $\sim200$~fs, long-term activation of the nanodroplets lasting up to a few nanoseconds is observed. Molecular dynamics simulations suggest that the short-term activation is caused by the injection of seed electrons into the droplets by XUV photoemission. Long-term activation appears due to electrons remaining loosely bound to photoions which form stable `snowball' structures in the droplets. Thus, we show that XUV irradiation can induce long-lasting changes of the strong-field optical properties of nanoparticles, potentially opening new routes to controlling avalanche-ionization phenomena in nanostructures and condensed-phase systems.


\end{abstract}

%
%
%
%
%

\section{Introduction}
Laser-induced nanoplasmas from clusters and nanoparticles are intriguing transient states of matter featuring extraordinary properties. Nanoplasmas are capable to efficiently absorb laser light and to convert it into energetic highly-charged ions, fast electrons and radiation covering the entire electromagnetic spectrum up to X-rays~\cite{saalmann_mechanisms_2006,fennel_laser-driven_2010}. Laser-induced nanoplasmas and microplasmas can potentially be used as compact accelerators for charged and neutral particles~\cite{fennel_laser-driven_2010,rajeev_compact_2013,KlimesovaPRA} and for generating pulsed extreme-ultraviolet (XUV) and x-ray radiation~\cite{kundu_harmonic_2007,vampa_linking_2015,gnodtke_ionization_2009,gorkhover_femtosecond_2016,masim_au_2016}.

While the ionization dynamics of clusters and nanodroplets induced by intense near-infrared (NIR) pulses is fairly well understood~\cite{saalmann_mechanisms_2006,fennel_laser-driven_2010,heidenreich_simulations_2007,heidenreich_charging_2017}, a few aspects still remain unresolved. These include the ignition process of a nanoplasma which occurs on the ultrashort timescale of electron motion (femtoseconds), as well as processes occurring at much longer times when the nanoplasma expands and charges recombine (picoseconds to nanoseconds). The plasma avalanche ionization is usually initiated by strong-field tunnel ionization (TI) of a few atoms in the nanocluster; the created quasi-free electrons are driven by the laser field unleashing an avalanche of secondary electrons mainly by electron-impact ionization (EII) of the surrounding atoms. Apart from tunnel ionization, seed electrons have been created by XUV and x-ray photoionization~\cite{schutte_ionization_2016,schomas_ignition_2020,kumagai2018following}. In the expansion phase of the nanoplasma, correlated electronic decay processes involving quasifree electrons and highly excited atoms and ions can occur~\cite{schutte2015observation,oelze2017correlated,niozu2019electron,kelbg2020temporal}. The latter have been observed as characteristic features in electron spectra which are usually dominated by a smooth exponentially decaying energy dependence due to thermal electron emission~\cite{schutte2014rare,medina_single-shot_2021}.
In all experiments reported so far, the regimes of ignition and expansion of the nanoplasma occur on very different timescales (femtoseconds vs. picoseconds up to nanoseconds, respectively). Here we present a case where laser-driven avalanche ionization of He nanodroplets is enhanced even after nanosecond delay times.

Helium nanodroplets are particularly well suited target systems for studying avalanche ionization and the ensuing nanoplasma dynamics. Due to their simple electronic structure and the large spacing between atomic levels, electron spectra are relatively easy to interpret~\cite{kelbg2019auger}. The extremely high ionization energy and its unique superfluid nature make He nanodroplets an ideal model system for studying the ionization dynamics of heterogeneous nanostructures. For example, by adding a few xenon atoms forming a cluster in the core of the droplet, the intensity threshold for avalanche ionization is drastically reduced~\cite{mikaberidze_laser-driven_2009,krishnan_dopant-induced_2011}. In contrast, alkali-metal clusters residing at the droplet surface turned out to be very inefficient in igniting avalanche ionization despite their extremely low ionization energy~\cite{heidenreich_efficiency_2016}. 

Here we exploit the properties of He nanodroplets to form stable He complexes around ions embedded in their interior, termed `snowballs'~\cite{atkins1959ions,Johnson:1972, gonzalez2020solvation}. Due to higher binding energy of an ion to the surrounding neutral He atoms in a He droplet (compared to the interactions between neutral He atoms), discrete shells of He atoms form around the ion whose density can exceed that of solid He. Stable, highly charged droplets containing snowballs have recently been studied by mass spectrometry and by means of x-ray diffractive imaging~\cite{laimer2019highly,feinberg2022x}. It should be noted that the formation of solid-like shells of atoms or molecules around positive ions by electrostriction is a general phenomenon occurring in all rare gases and non-polar dielectric liquids~\cite{hiraoka1990stability,schmidt1999structure}. 

In contrast to positively charged ions, electrons tend to form extended void bubbles inside He droplets or to reside in metastable states at the droplet surface~\cite{rosenblit1995dynamics,farnik1998differences}. In the context of nanoplasma generation by laser-induced avalanche ionization, both the formation of snowballs around ions and the attachment of electrons to the nanodroplets may facilitate avalanche ionization; The presence of ions in the vicinity of He atoms tends to lower their ionization threshold by 10-17~eV~\cite{Heidenreich_2017}, and the quasi-free electrons present in highly charged nanodroplets act as seeds for laser-driven impact ionization. Here we present systematic time-resolved experiments of large He droplets irradiated by weak XUV pump pulses and strong NIR probe pulses. In addition to an enhancement of avalanche ionization of the droplets at short pump-probe delays on the timescale of 200~fs, we observe enhanced avalanche-ionization rates at long pump-probe delays up to nanoseconds. Dedicated model simulations suggest that this long-lasting activation of the droplets by XUV irradiation is due to the simultaneous presence of snowballs and quasi-free electrons in the He droplets.

\section{Experimental setup}
\begin{figure}
	\center
	\includegraphics[width=1.0\columnwidth]{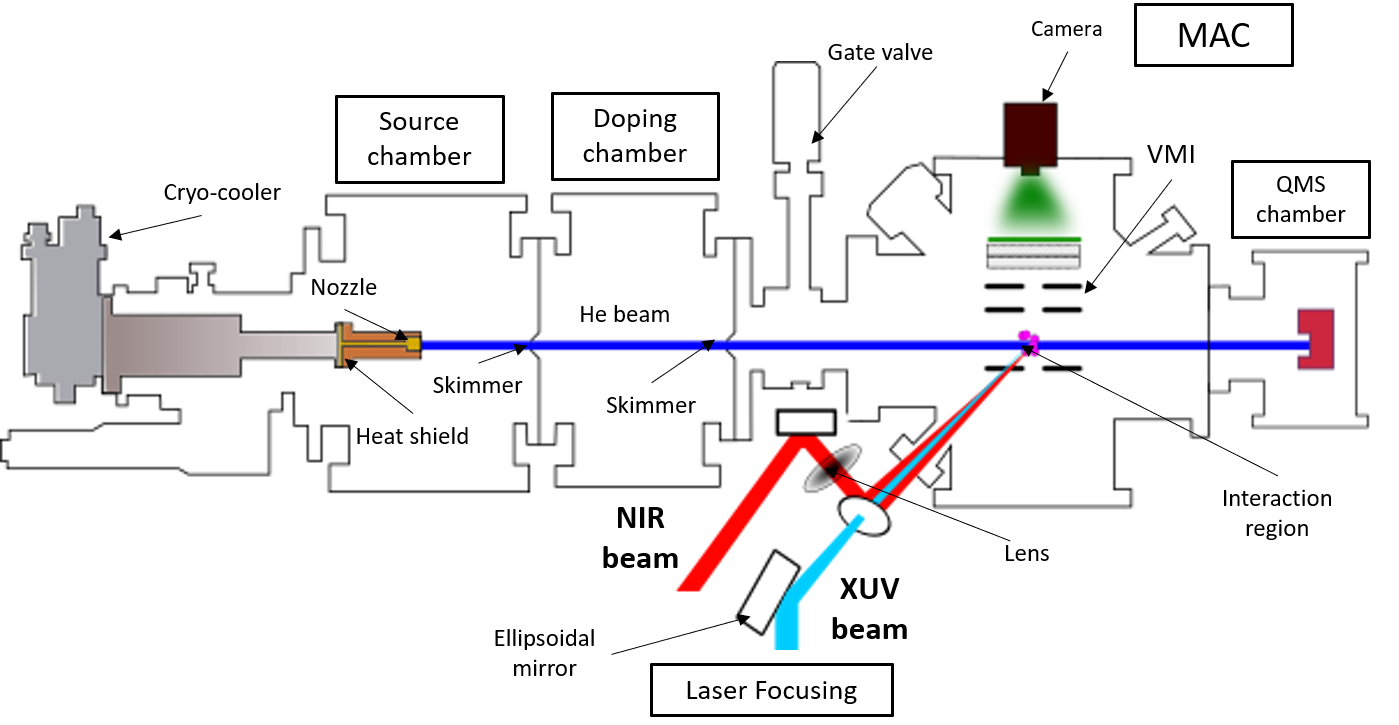}
	\caption{Schematic representation of the experimental setup. The He nanodroplet beam propagates from left to right along the horizontal blue line. The NIR and XUV beams are superimposed and focused into the interaction region of a velocity-map imaging spectrometer inside the MAC chamber. \label{fig:setup} }
\end{figure}

The experiments were performed using the \textit{Multipurpose end-station for Atomic, molecular and optical sciences and Coherent diffraction imaging} (MAC) at the Extreme Light Infrastructure (ELI beamlines) user facility in Doln\'{i} B\v{r}e\v{z}any near Prague~\cite{klimesova_multipurpose_2021}. The experimental setup is outlined in Fig.~\ref{fig:setup}. A beam of He nanodroplets was generated by a continuous supersonic expansion using ultrapure He at a pressure of 50~bar out of a cryogenic nozzle. The average droplet sizes were in the range of $\langle N\rangle =9\times 10^4$ - $7\times 10^5$ He atoms per droplet. Droplet sizes were determined from titration measurements, see the Supplemental Material (SM)~\cite{gomez_sizes_2011}. The collinearly propagating XUV and NIR laser beams were focused into the He droplet beam inside the interaction region of an electron velocity map imaging (VMI) spectrometer~\cite{klimesova_multipurpose_2021}. At the estimated number density of He droplets in the interaction region of about $10^6~$cm$^{-3}$, the maximum rate of detected avalanche ionization events per laser shot was about 8\,\%. As a characteristic feature of avalanche ionization of clusters and nanodroplets by intense NIR pulses, the signals of individual hits are detected as bright round distributions at the electron detector (microchannel plate and phosphor screen)~\cite{medina_single-shot_2021}. Since a single avalanche-ionized He droplet creates a large number of electrons, the electron detector was operated at low gain to ensure approximately linear signal response despite the widely fluctuating signal intensities. The images are analyzed by integrating their total brightness and by determining the widths of the radial profiles of the bright spots from gaussian fits. From the widths we infer the average electron energies $E_e$. The VMI spectrometer is calibrated by recording photoelectrons emitted from rare gas atoms injected directly into the spectrometer chamber and irradiated with XUV pulses alone. 

The XUV beamline~\cite{hort2019high} was driven by a commercial laser system (Legend Elite Duo from Coherent) operating at 1~kHz with a pulse length $\leq35$~fs. It provides pulses with up to 12~mJ energy at a central wavelength of 795~nm. About 10\,\% of the beam was redirected to a delay stage that has a total travel range of 1~m corresponding to about 6~ns of variable delay. The more intense part of the beam was focused into a high-harmonic generation (HHG) unit consisting of a pulsed gas expansion out of an Even-Lavie-type pulsed valve running at a repetition rate of 500~Hz. The HHG beamline was operated with Kr gas and was optimized to achieve highest intensity around the $19^{th}$ harmonic (HH19) or $21^\mathrm{st}$ harmonics (HH21) at photon energies $29.6~$eV and $32.8$~eV, respectively. One of the two harmonics was selected using a grating monochromator. The photon flux was measured by an energy-calibrated XUV-sensitive photodiode yielding a photon flux of about $10^7$ photons/pulse for HH19 and $6\times10^6$ photons/pulse for HH21 in the interaction region. A FWHM energy bandwidth of the XUV pulse of about 300~meV was determined by a grating spectrometer located next to the monochromator.
 
The NIR laser beam was focused by a lens with a focal length of 200~mm and co-linearly superimposed onto the XUV beam using a plane mirror with a centered hole. The XUV beam was focused by an ellipsoidal mirror with a focal length of 500~mm to a focal spot size (FWHM) of $35\times 50~\mu$m$^2$ yielding an intensity of $5\times10^{7}$~Wcm$^{-2}$. In the presented experiments, the NIR pulses were attenuated to 230~$\mu$J or less. With a focal spot size of $15\times 20~\mu$m$^{2}$, the NIR peak intensity was $3\times10^{14}$~Wcm$^{-2}$. The FWHM pulse duration of the NIR pulses was 150~fs after beam transport in vacuum (4~m of travel) and through air (20~m) to the MAC chamber. The XUV pulses have an estimated duration of 50~fs after the monochromator~\cite{klimesova_multipurpose_2021}. The spatial overlap of the NIR and XUV pulses was adjusted by imaging the two beams on a YAG:Ce fluorescence screen inserted into the focal plane. The temporal overlap of the pulses was determined by measuring transient Kr$^{2+}$ ion yields from Kr atoms injected into the spectrometer~\cite{drescher_time-diagnostics_2010,roos_2022}. The zero delay, $\Delta t=0$, was determined with a precision of 10~fs. The exposure time of the camera recording electron VMI's was 10~ms thereby averaging over 5 consecutive XUV and, correspondingly, 10 NIR pulses. For each data point of the pump-probe curves shown below, 1000 VMI's were recorded.

\begin{figure}
    \centering
    \includegraphics[width=0.6\columnwidth]{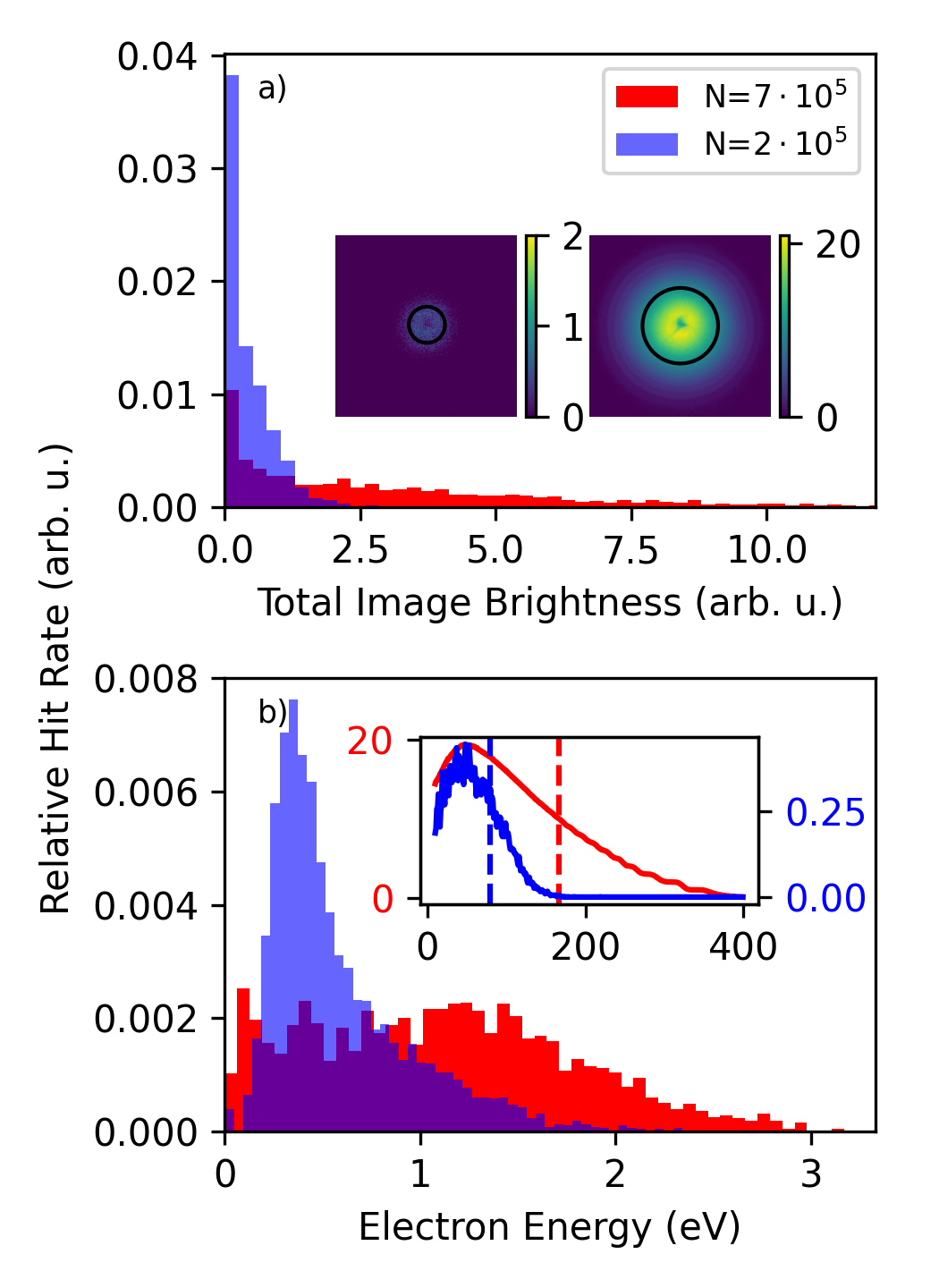}
    \caption{a) Histograms of the total brightness of electron velocity-map images recorded with XUV pump and NIR probe pulses at two He droplet sizes, $\langle N\rangle = 2\times 10^5$ (blue bars) and $\langle N\rangle = 7\times 10^5$ (red bars). The inset shows examples of electron images with low brightness (left) and with high brightness (right). The circles depict the range where the electron intensity has dropped to half of the peak intensity as determined from gaussian fits. b) Histograms of electron energies inferred from the images. The inset shows the radial intensity profiles of the two images in the inset in a). The dashed vertical lines indicate the radii of the circles in the insets in a).
    \label{fig:histo}
    }
\end{figure}

\section{Experimental XUV-pump NIR-probe dynamics}
Electron signals created by avalanche ionization of rare-gas clusters and He nanodroplets are subjected to large shot-to-shot fluctuations. This is mainly due to the large variation of sizes of the individual droplets following a broad droplet-size distribution~\cite{knuth1999average}. In addition, the laser intensity seen by each droplet varies greatly as each droplet is hit by a laser pulse at a different position within the laser
focus~\cite{medina_single-shot_2021}. Fig.~\ref{fig:histo} a) shows histograms of the integrated image brightness which is proportional to the total number of detected electrons in one hit. Histograms of the characteristic electron energies $E_e$ are shown in panel b). The two distributions in each panel correspond to He droplets of different mean sizes, $N\approx 2\times 10^5$ (blue) and $N\approx 7\times 10^5$ He atoms per droplet (red). About $4\times 10^4$ images were analyzed for each histogram. In these data, the delay between XUV and NIR pulses was varied in the delay range 1.5~ps - 5~ns where no systematic variations of the data were observed. Examples of individual electron VMI's are shown as insets in Fig.~\ref{fig:histo}~a) and the corresponding radial profiles are shown as inset in b). From the circles with radius $R$ placed around each electron distribution at one half of the peak intensity as obtained from two-dimensional gaussian fits [vertical dashed lines in the inset in b)] we determine the characteristic electron energy, $E_e=kR^2$~\cite{medina_single-shot_2021}. The factor $k=7.22$~eV/pixel$^2$ is obtained from calibration measurements where Kr atoms were directly photoionized by the XUV pulses.

While the histograms of total image brightness are peaked at very low values and smoothly fall off toward higher brightness, the histograms of electron energy show a maximum at finite $E_e$. Clearly, larger He droplets produce on average brighter signals covering larger areas on the electron detector due to higher $E_e$. In fact, a clear correlation of the number of electrons (image brightness) and electron energy $E_e$ (deduced from radius) is observed on the level of individual events, as discussed in a previous study~\cite{medina_single-shot_2021}. 

\begin{figure}
    \centering
    \includegraphics[width=1.0\columnwidth]{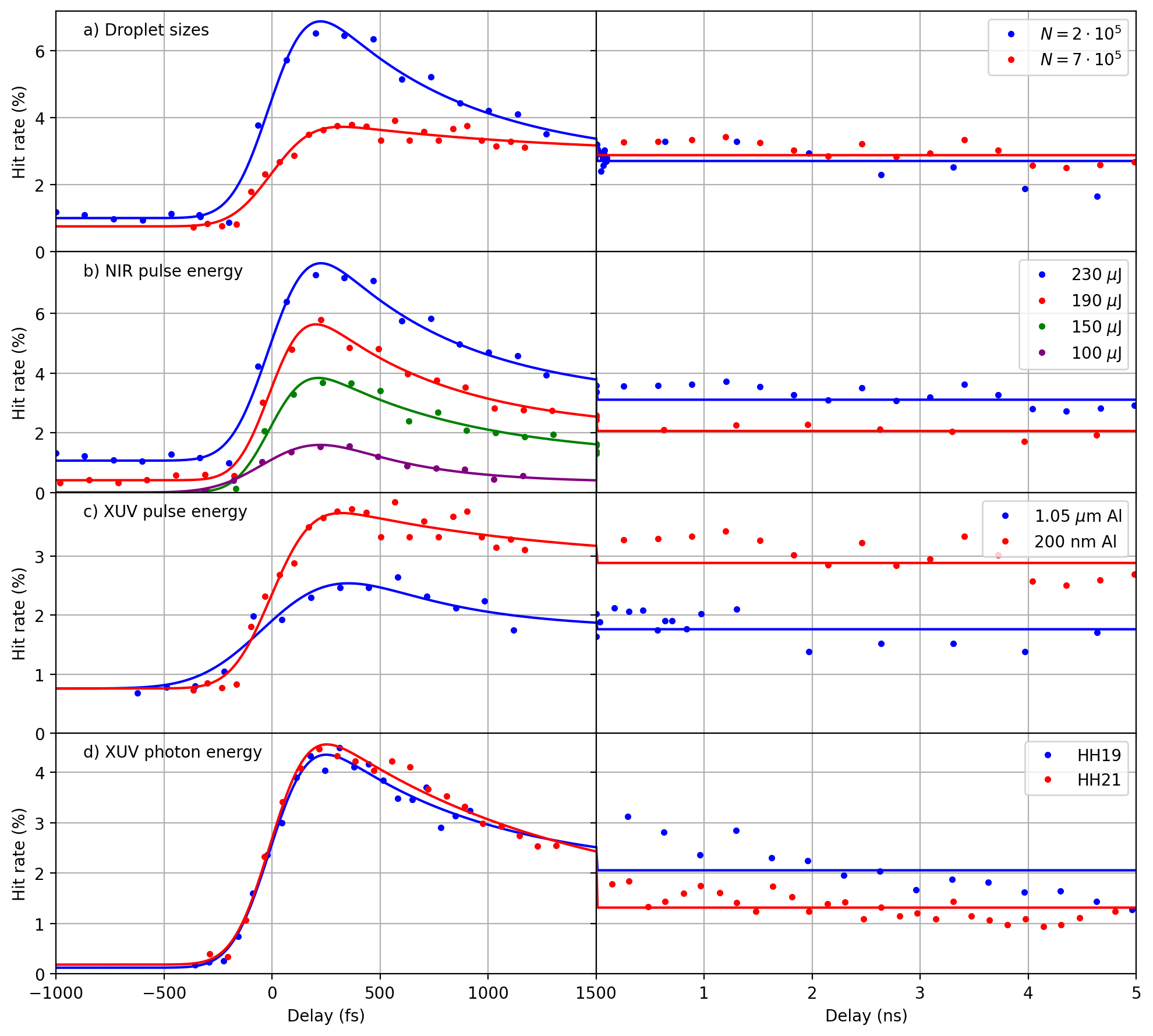}
    \caption{Transient hit rates at different experimental parameters as a function of XUV pump and NIR probe delays. The hit rate is defined as the percentage of He droplet avalanche-ionization events per XUV-NIR pulse pair. In a), b), c), and d) the He droplet size, the NIR pulse energy, the XUV pulse intensity, and the harmonic order were varied, respectively. In a)-c) HH21 was used to generate the XUV pump pulse. In b) and d) the mean droplet size was $\langle N\rangle = 2\times 10^5$, in c) it was $\langle N\rangle = 7\times 10^5$. In a) and c) the probe pulse energy was $230~\mu$J, in d) it was $160~\mu$J. \label{fig:expPP}
    }
\end{figure}

The most pronounced effect of the XUV pulses on He-droplet avalanche ionization driven by NIR pulses is observed on the detection rate of images containing nanoplasma electrons in proportion to the XUV-pulse repetition rate (hit rate). The NIR pulses are attenuated to a level that the hit rate from the NIR pulses alone, which is equal to the hit rate at $\Delta t\ll 0$, is low in the sense that when XUV pulses are added before the NIR pulses the hit rate is enhanced by at least a factor 10 at optimum delay ($\Delta t=200$~fs). Note that the hit rate at $\Delta t\ll 0$ is enhanced by factor 2 as every other NIR pulse interacts with the He droplets in the absence of a XUV pulse. Fig.~\ref{fig:expPP} shows the hit rate in proportion to all recorded images as a function of XUV-pump and NIR-probe delay $\Delta t$ for short delays $\Delta t \leq 1.5$~ps (left panel) and for much longer delays $\Delta t=0$ to 5~ns (right panel). A common feature of these pump-probe curves is a steep rising edge centered around $\Delta t=0$ followed by a maximum around $\Delta t=200~$fs. For longer delays, the hit rate drops again on the time scale of $\sim 1~$ps. The surprising new observation is that the hit rate remains significantly enhanced even if the NIR pulse is delayed by several nanoseconds with respect to the XUV pulse. At such long delays after XUV activation, free electrons were assumed to be emitted from the droplets or to recombine with their parent ions. A long-lasting enhancement of the hit rate indicates that XUV activation of the He droplets for NIR-driven avalanching persists for extended time periods.

Based on these observations, we use a fit model of the nanoplasma pump-probe dynamics given essentially by the sum of two terms; One describes a transient enhancement of the ignition rate near $t=0$ due to resonant coupling of NIR probe pulse to the XUV-activated He nanodroplets, $S_\mathrm{short}(t)=A\,\Theta(t)\exp(-t/\tau)$. Similar dynamics has previously been observed in NIR and soft X-ray-activated  He nanodroplets~\cite{Krishnan_2012,schomas_ignition_2020}. The second term accounts for long-lasting activation by the XUV pump pulse which we assume to be constant for $t\gg 0$, $S_\mathrm{long}(t)=B\,\Theta(t)$. Here $\Theta(t)$ is the Heaviside step function. As the temporal resolution in pump-probe measurements is limited by the cross-correlation of the pump and probe pulses $M(t) = \exp\left( -t^2/(2\sigma^2\right)/(\sqrt{2\pi}\sigma)
$, the resulting model function is given by the convolution of $S_\mathrm{short}(t)+S_\mathrm{long}(t)+C$ and $M(t)$ resulting in~\cite{laforge2022relaxation}
\[
S(t) = A'\,\exp\left(-\frac{t}{\tau}\right)\textrm{erfc}\left(\frac{\sigma}{\sqrt{2}\tau} - \frac{t}{\sqrt{2}\sigma}\right) + B'\,\textrm{erfc}{\frac{-t}{\sqrt{2}\sigma}}+C.
\]
The standard deviation $\sigma$ is related to the FWHM of the cross-correlation function by $\sigma = \textrm{FWHM}/\sqrt{8\ln 2}$. Additionally, a constant offset $C$ is added to this function to account for the finite hit rate induced by the NIR pulse before the He droplets are activated by the XUV pulses, \textit{i.\,e.} at $t<0$. Note that this value should be divided by 2 to account for the twice higher repetition rate of the NIR pulses compared to the XUV pulses. 

When inspecting the pump-probe curves recorded for different experimental conditions -- variable average He droplet sizes [Fig.~\ref{fig:expPP} a)], variable NIR probe-pulse energy using an adjustable neutral-density filter (b), variable XUV pump-pulse energy achieved by inserting an additional metallic filter into the beam (c), and for two harmonics HH19 and HH21 (d) -- we note the following trends: \\
(i) The width of the rising edge around $\Delta t=0$ is nearly constant for all conditions, $\sigma=145\pm 5~$fs, corresponding to 330~fs FWHM. This is about twice the expected width of the cross-correlation of the pump and probe pulses. Thus, the characteristic time for XUV-NIR ignition dynamics up to reaching the maximum ignition probability is about $330/\sqrt{2}$~fs = 230~fs.\\
(ii) The hit rate around $\Delta t=200~$fs is more strongly enhanced for smaller droplets $N=2\times 10^5$ as compared to droplets of average size $N=7\times 10^5$; However, the hit rate at long delays remains equal for both droplet sizes. \\
(iii) The exponential short-time decay constant $\tau$ is found to depend only on the He droplet size, $\tau = 650\pm 40$~fs for $\langle N\rangle = 2\times 10^5$ and $\tau = 1050\pm 50$~fs for $\langle N\rangle = 7\times 10^5$. This trend was previously observed for different ignition schemes~\cite{Krishnan_2012,schomas_ignition_2020}. Larger systems tend to expand more slowly as the core remains quasi-neutral and electrons and ion remain confined in the system for longer periods of time inner (incomplete outer ionization)~\cite{last_quasiresonance_1999,last2004electron}.\\
(iv) Increasing the NIR probe-pulse energy from $100$ to $230~\mu$J enhances the hit rate in the full range of $\Delta t$; At negative delays, \textit{i.\,e.} when the NIR pulse interacts first, avalanche ionization is induced by the NIR pulse alone and the XUV has no measurable effect; Clearly, more intense NIR pulses lead to an enhanced hit rate. At short delay $\Delta t\approx 200$~fs, the rise of the maximum ignition rate is even more pronounced. The long-term hit rate is also systematically enhanced.\\
(v) Attenuation of the XUV pump pulse mainly reduces the hit rate in the full range of positive delays.\\
(vi) No significant difference in the hit rate is seen for the two photon energies of the XUV pulses, $29.6~$eV (HH19) and $32.8$~eV (HH21).

\begin{figure}
    \centering
    \includegraphics[width=0.6\columnwidth]{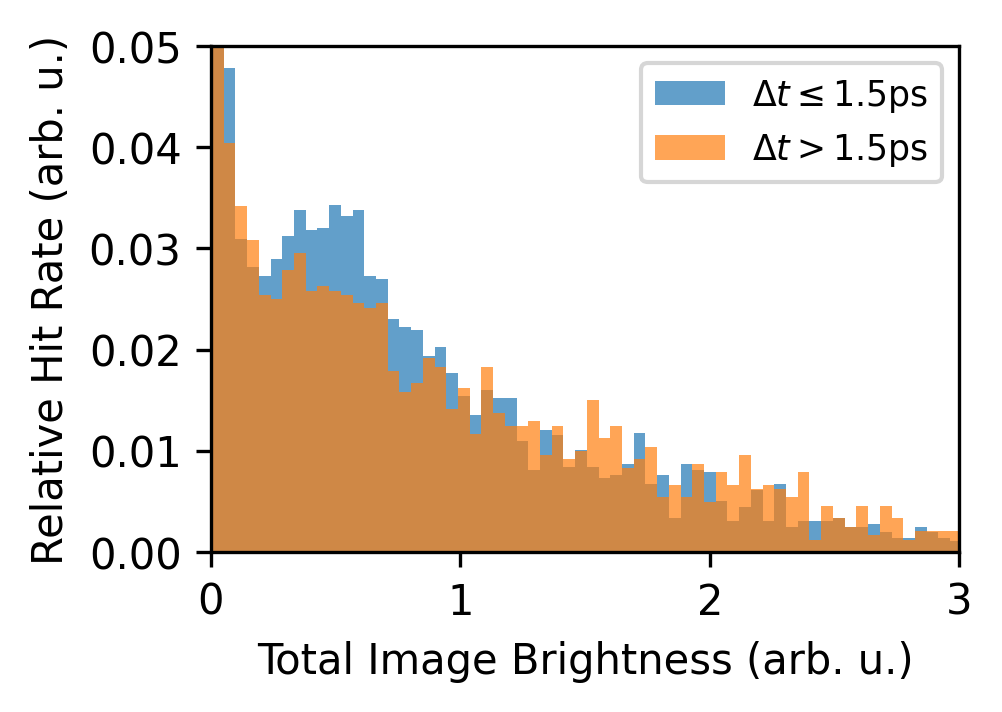}
    \caption{Histograms of integrated image brightness of electron VMI's recorded at short XUV-pump and NIR-probe delays $\Delta t\leq 1.5~$ps and at $\Delta t > 1.5~$ps for harmonic HH21, $\langle N\rangle = 2\times 10^5$ and $200~\mu$J probe pulse energy. \label{fig:histoPP}
    }
\end{figure}

While the hit rate shows a strong pump-probe contrast and the XUV pulse fully controls the probability of avalanche ionization under certain conditions (see e.~g. the curves in Fig.~\ref{fig:expPP} b) at low-intensity), the structure of the nanoplasma electron VMI's only slightly depends on $\Delta t$. Fig.~\ref{fig:histoPP} shows histograms of the integrated brightness of electron images recorded using harmonic HH21 at short pump-probe delays $\Delta t\leq 1.5~$ps (blue) and at long delays $\Delta t \gg 1.5~$ps (yellow). The main difference is that at $\Delta t\leq 1.5~$ps, the distribution features a local maximum in the range of low brightness around 0.5 arb.~u. which is less pronounced in the distribution for $\Delta t > 1.5~$ps. Nanoplasma events with low total electron yield mostly originate from small droplets which are most abundant in the droplet beam~\cite{knuth1999average}. This finding is in line with the enhancement of the hit rate of small droplets at short delays as observed for different mean droplet sizes, see Fig.~\ref{fig:expPP} a). Activation of small nanodroplets by XUV ionization appears to be more of a transient phenomenon; Photoelectrons and ions rapidly expand and mostly leave the droplets within $\sim1$~ps. In contrast, large He droplets tend to trap ions and electrons in stable and metastable states, respectively, thereby rendering the droplets susceptible for laser-induced avalanche ionization over longer periods of time. The observation that the ignition rate does not appear to be proportional to the XUV intensity, which is varied by nearly a factor 10 in Fig.~\ref{fig:expPP} c), may indicate that the droplet sizes used in this experiment can sustain only a limited number of snowball complexes in stable states; Higher XUV ionization rates would induce the ejection of ions from the droplets. Note that the droplet sizes in this experiment are close to the minimum size needed to support multiple charges, $N=10^5$~\cite{laimer2019highly}.


Efficient ignition of NIR-induced nanoplasmas by high-harmonic XUV pulses has previously been demonstrated for argon clusters~\cite{schutte_ionization_2016}. However, no pump-probe dynamics were reported. In a previous experiment, we used soft x-ray pulses from a free-electron laser (FEL) to irradiate He droplets of size $\sim10^4$ He atoms per droplet doped with heavier rare-gas atoms~\cite{schomas_ignition_2020}. We observed a similar increase of the NIR-induced avalanche ionization rate as in the current study within the delay range of overlapping XUV and NIR pulses. However, in contrast to this study, the avalanching rate dropped again to nearly zero at XUV-NIR delays $\Delta t\leq 2$~ps; No long-lasting ionization was observed. It should be noted that in that FEL experiment the droplets were ignited mainly by inner-shell ionization of the dopant clusters due to the low absorption cross-section of He at the higher photon energy (250~eV). The dopant clusters were multiply ionized as the XUV pulse intensity was much higher ($3\times 10^{13}~$Wcm$^{-2}$) than in the present experiment ($5\times 10^7~$Wcm$^{-2}$). 

\begin{figure}
    \centering
    \includegraphics[width=1.0\columnwidth]{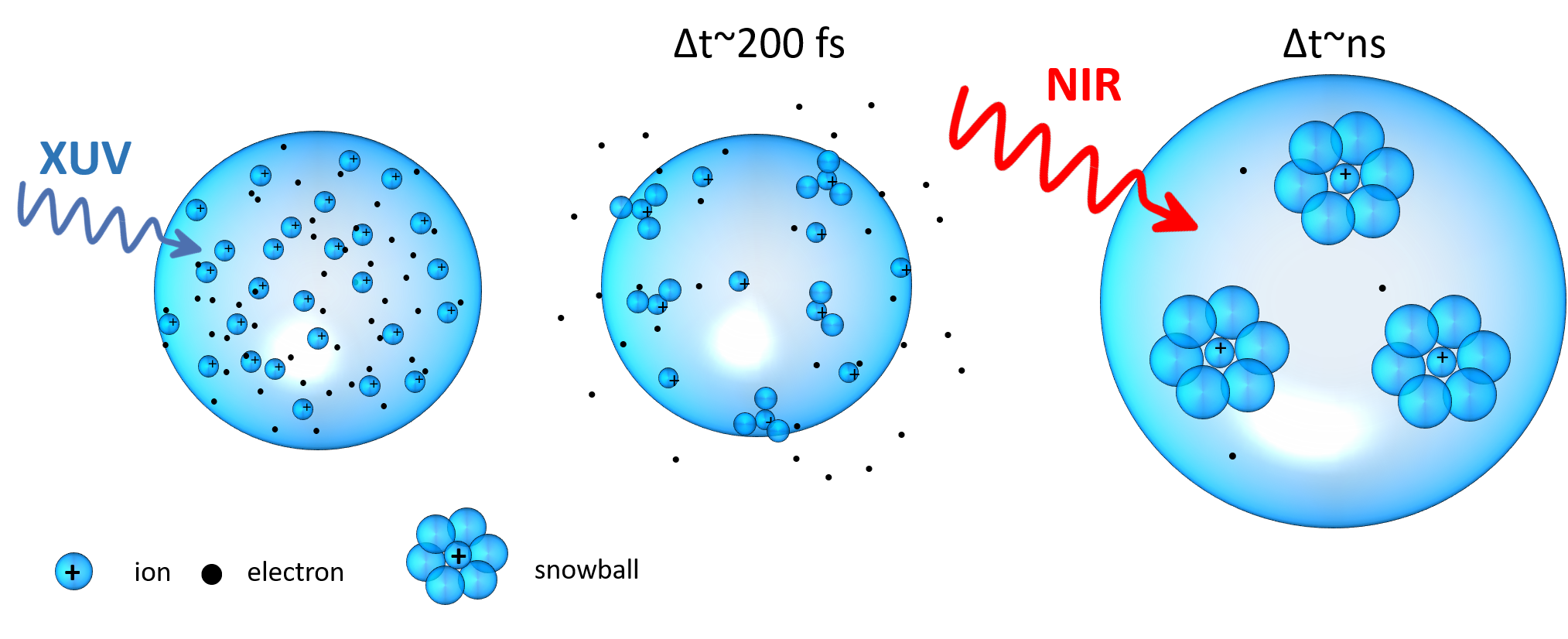}
    \caption{Schematic representation of the suggested dynamics occurring in He nanodroplets irradiated by an XUV pulse followed by a NIR pulse after some time delay $\Delta t$. The regular clusters of blue spheres around ``+'' depict snowball complexes forming around He$^+$ cations. The black dots depict quasi-free electrons. \label{fig:cartoon}
    }
\end{figure}

Thus, in the present experiment, a new XUV activation process must be active in pure He droplets at long delays. Here we argue that photoions forming stable snowball complexes in He droplets as well as quasi-free electrons trapped in the charged droplets are the origin of the long-term activation of the droplets. Snowballs facilitate NIR avalanche ionization by the lowering of the ionization energy of He atoms bound to ions, and quasi-free electrons act as seeds for NIR-induced avalanching. A schematic view of the proposed dynamics is given in Fig.~\ref{fig:cartoon}. 

\section{Molecular dynamics simulations}
To get more detailed insights into both the short-delay pump-probe dynamics and the long-term ionization process, we performed classical molecular dynamics (MD) simulations. The general features of the MD simulation method for the interaction of a cluster with the electric and magnetic field of a linearly polarized NIR Gaussian laser pulse were described previously~\cite{heidenreich_simulations_2007,heidenreich_charging_2017,schomas_ignition_2020,heidenreich_efficiency_2016,heidenreich_ion_2012}. In short, all atoms and nanoplasma electrons are treated classically, starting with a cluster of neutral atoms. Electrons enter the MD simulation upon XUV photoionization or when the criteria for TI, classical barrier suppression ionization (BSI) or EII are met. Conventionally, all channels by which atoms or molecules of a cluster or droplet are ionized, such as TI, BSI, EII and photoionization, are subsumed by the term inner ionization; The stripping of nanoplasma electrons from the cluster or droplet nuclear framework is termed outer ionization~\cite{last_quasiresonance_1999}. The criteria for TI, BSI and EII are checked at each atom at every MD time step, using the local electric field at the atoms as the sum of the laser electric field and the contributions from all ions and electrons of the cluster. Instantaneous TI probabilities are calculated by the Ammosov-Delone-Krainov formula~\cite{ammosov_tunnel_1986}, EII cross sections by the Lotz formula~\cite{lotz_subshell_1968}, taking the ionization energy with respect to the local atomic Coulomb barrier in the cluster~\cite{fennel_highly_2007}. Interactions between ions are described by Coulomb potentials, electron-ion and electron-electron interactions by smoothed Coulomb potentials~\cite{Ditmire:1998,last_quasiresonance_1999}. Interactions involving neutral atoms are disregarded except for a Pauli repulsive potential between electrons and neutral He atoms. Chemical bonding involving electronic open-shell atoms is disregarded, so that the formation of He$_n^+$ complexes is not accounted for. Such He$_n^+$ complexes, which are the subject of this paper, will be introduced in an \textit{ad hoc} manner as described in the following section.

To simulate the transient NIR avalanching rates on the few-picosecond time scale, the XUV pulse (photon energy 32.8~eV) is not described explicitly but mimicked by creating $N_\mathrm{pump}$ He$^+$-electron pairs randomly distributed in the droplet as part of the initial conditions of the trajectories, that is to say, as instantaneous ionizations. All photoelectrons are placed at a distance of 2.5~\AA~in random directions from their parent atoms. The initial photoelectron kinetic energy is corrected for the interaction with the parent ion and with the droplet environment. That is to say, with the initial distance of 2.5~\AA~from their parent ions and the corresponding residual kinetic energies, the photoelectrons are given plenty of opportunities to collide with other atoms on their path through the droplet so that an equilibrium of electrons leaving the droplet or being retained by the droplet can establish in a natural way. In the experiment, the NIR probe-pulse peak intensity in the center of the focus was $2$ - $3\times10^{14}$~Wcm$^{-2}$ (Gaussian intensity FWHM $t_\mathrm{FWHM} = 150$~fs which corresponds to a FWHM $\tau = 212$~fs of the Gaussian electric field envelope). Since focal averaging~\cite{heidenreich_kinetic_2011} is not considered in the present simulations, we choose a pulse peak intensity $I_M = 10^{14}$~Wcm$^{-2}$ which represents some average in the inner region of the NIR focal spot. Test simulations for a He$_{10149}$ droplet at short delays of the NIR pulse arrival $\Delta t = 200$~fs reveal that the pump-probe signals (He ion and electron yields) begin to emerge at a number $N_\mathrm{pump}\approx20$ XUV photoionizations per droplet in part of the trajectories, and for $N_\mathrm{pump}\approx30$ the He ionization avalanche takes place in every trajectory. For the simulation of the pump-probe signal curve $N_\mathrm{pump}=40$ and $55$ is chosen.

Given the experimental XUV pump-pulse characteristics ($10^7$ photons per pulse, $35\times 50~\mu$m$^2$ spot size) and the photoionization cross sections of He at a photon energy of 32.8~eV (4.57~Mb)~\cite{samson_precision_1994}, the photoionization probability of a single He atom per pulse is $3.3\times10^{-6}$. 
This implies that measurable pump-probe signals are generated by He droplets of size $\gtrsim 6\times10^6$ atoms containing at least $N_\mathrm{pump}\approx20$ XUV photoionizations according to the simulation. 
Given the experimental average droplet sizes ($2$-$7\times10^5$ He atoms), these large droplets, present in the tail of the droplet size distribution, contribute most to the detected pump-probe signal~\cite{knuth1999average}. Unfortunately, droplets of this size exceed our computational capabilities; Therefore we restrict our simulations to droplets sizes of $10^4$ atoms. The initial structure of the He nanodroplet before irradiation is assumed to be a fcc lattice with an interatomic He-He distance of 3.6~\AA~\cite{peterka_photoionization_2007}. 

The efficiency of an avalanche ionization event in a single trajectory can be quantified by the average He charge
\begin{equation}
\centering
    \label{eq:kappa}
    \langle q_\mathrm{He}\rangle=\frac{N_\mathrm{He^+} + 2N_\mathrm{He^{2+}}}{N_\mathrm{He} + N_\mathrm{He^+} + N_\mathrm{He^{2+}}}.
\end{equation}
Here, $N_\mathrm{He}$, $N_\mathrm{He^+}$ and $N_\mathrm{He^{2+}}$ are the numbers of He, He$^+$ and He$^{2+}$ species at the end of a trajectory. Although the trajectory simulations in principle allow for classical three-body recombination (TBR), the temporal length of the trajectories is too short for TBR to be completed. Accordingly, we do not check for TBR and $N_\mathrm{He},~N_\mathrm{He^+} $ and $N_\mathrm{He^{2+}}$ are the populations of bare atom charges as obtained by inner ionization.

A simulated average He charge $\langle q_\mathrm{He} \rangle$ of a trajectory corresponds to the number of electrons of a single droplet in the experiment. In the experiment the ignition probability is given by the fraction of droplets in which a given threshold of electron counts is exceeded. To directly relate our simulations to the experimental nanoplasma hit rate, we determine the ignition probability by evaluating the proportion of trajectories for which $\langle q_\mathrm{He}\rangle$ transcends a given threshold. To define a suitable threshold for $\langle q_\mathrm{He} \rangle$, we assume that the ionization avalanche starts at a single location in a droplet and propagates from there, so that a small simulated droplet can be viewed as an excerpt of a large droplet in the experiment. Accordingly, one should anticipate that the ionization avalanche has a chance to propagate through a large droplet only if at least the vast majority of He atoms of a small droplet is ionized at the end of a trajectory. Here we choose a threshold value of $\langle q_\mathrm{He} \rangle = 1$ for which resonant avalanche ionization was observed. In many cases, $\langle q_\mathrm{He} \rangle = 1$ was found to be realized when 90~\% of the He atoms are singly ionized and the rest remains neutral or is doubly ionized. It turned out that the exact value of the threshold of $\langle q_\mathrm{He}\rangle$ is uncritical in the range of 0.2-1.8 for the pulse parameters $I_M = 10^{14}$~Wcm$^{-2}$, $\tau = 212$~fs. When avalanche ionization occurs in the simulation, $\langle q_\mathrm{He}\rangle$ reaches nearly 2. In the absence of ignition, $\langle q_\mathrm{He}\rangle\approx0$. Deviations from this bimodality are only observed for negative time delays (rising edge of the pump-probe curve), when the effective part of the NIR pulse is truncated and consequently leaves many He atoms singly ionized.

For each pump-probe delay time, a set of $N_\mathrm{traject}$ trajectories with different initial conditions (slightly different initial atomic coordinates, different random locations of the $N_\mathrm{pump}$ He$^+$-electron pairs created by the pump pulse, a different initialization of the random number generator for TI) are simulated. The number $N_\mathrm{traject}$ of trajectories per set varies between 15 for high ignition probabilities (ignition of a He avalanche ionization in every trajectory, which is the case for short positive pump-probe delays $\Delta t$) and 100 for low ignition probabilities (ignition only in a fraction of the trajectories occurring for negative or large positive $\Delta t$). 

\section{Discussion}

\begin{figure}
	\center
	\includegraphics[width=0.7\columnwidth]{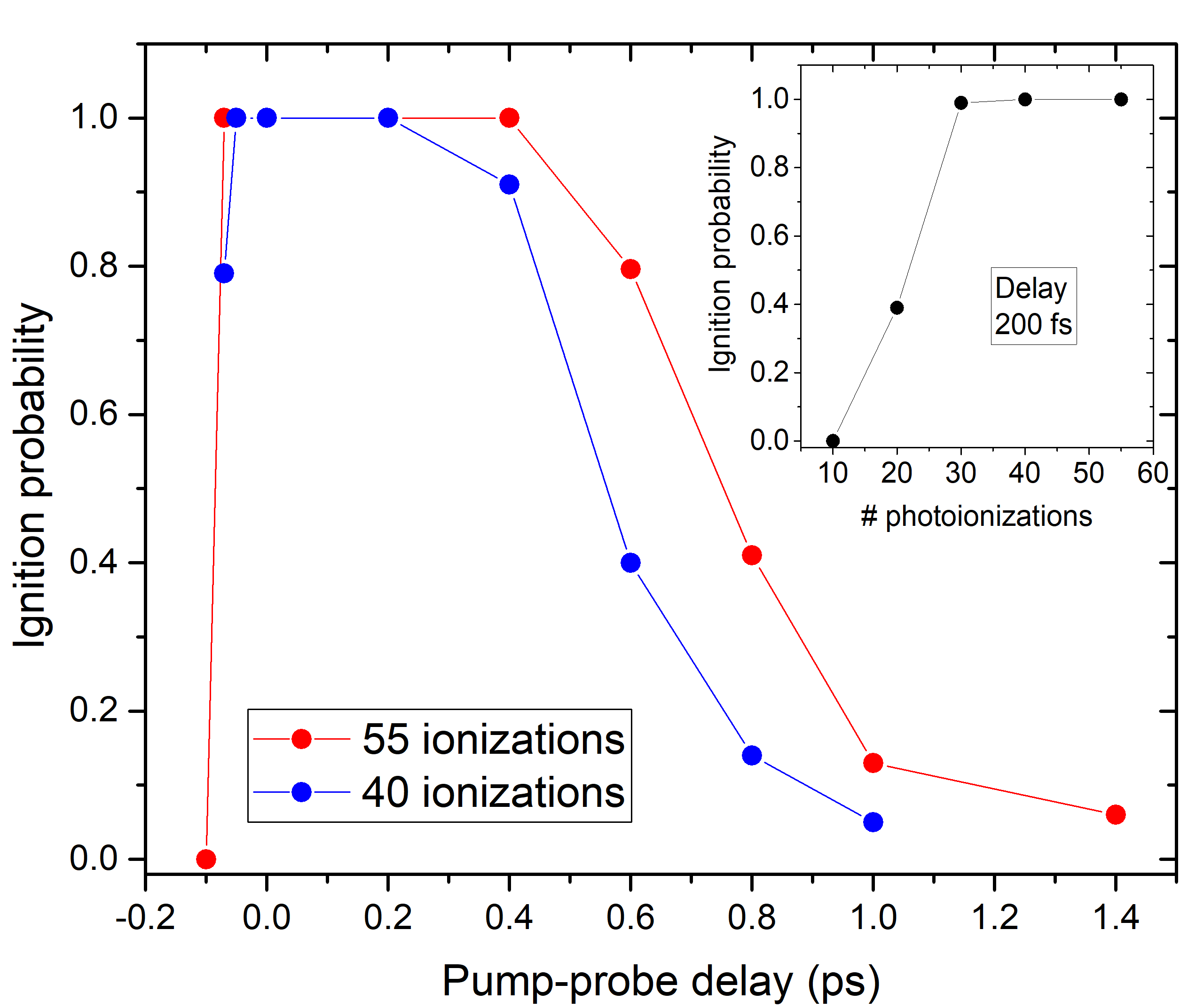}
	\caption{\label{fig:pssi} 
	Simulated pump-probe curves at short delays $\Delta t$ for different amounts of initial photoionizations in a He nanodroplet composed of $N=10,149$ He atoms. The inset shows the ignition probability at fixed delay $\Delta t = 200~$fs. 
	}
\end{figure}
Fig.~\ref{fig:pssi} shows the simulated ignition probability as a function of the pump-probe delay for 40 and 55 photoionized He atoms in one droplet of size $N=$10,149. In agreement with the experimental data, the ignition probability promptly rises around $\Delta t=0$ when the probe pulse acts on the XUV-photoionized He nanodroplet. In the simulation, nanodroplets subjected to 40 and 55 initial ionizations are fully ionized by the NIR pulse because partial recombination of electrons and ions in the later phase of the expansion is disregarded. The inset shows the ignition probability at fixed delay $\Delta t=200$~fs as a function of the number of XUV-induced photoionizations. Thus, it takes 30 photoionization events to fully avalanche ionize a He nanodroplet by every NIR pulse. At delays $\Delta t > 200$~fs, respectively $\Delta t > 400$~fs, the signal falls off again. Since the photoelectrons act as seeds for the NIR laser-driven EII avalanche, the falling edge reflects the escape of photoelectrons and He$^+$ ions from the droplets as the retaining potential created by the expanding photoions flattens out. Details of the electron and ion picosecond pump-probe dynamics are discussed in the SM. 

The time scale of the decrease of the ignition probability is in good agreement with the experimental results (Fig.~\ref{fig:expPP}). The somewhat more extended falling edge of the experimental pump-probe curves is likely the effect of averaging over the broad distribution of He droplet sizes and laser intensities, not taken into account in the simulations. However, in contrast to the experimental result, the simulated ignition probability drops nearly to zero at $\Delta t > 1$~ps, indicating that this version of the simulation does not accurately describe the long-term dynamics of XUV-pump, NIR-probe-ionized He nanodroplets. The main deficiency is likely the lack of He$^+$-He interactions which causes photoions to rapidly expand and leave the droplet even when only few ions are present and Coulomb repulsion is weak. In the real system at least some of the He$^+$ ions form He$_2^+$ and He$_3^+$ molecular ions. Subsequently, some of these molecular ions form snowball complexes by gathering a shell of neutral He ligand atoms within picoseconds~\cite{leal2014picosecond}. Since the interatomic distance between the surrounding neutral He ligand atoms and the central He$_2^+$ or He$_3^+$ ion is shortened from the bulk value of $3.6$ to $1.9~$\AA~\cite{knowles_structures_1996}, the Coulomb barriers at the ligand atoms are considerably reduced.
Approximating a He$_2^+$ ion core by two point charges of $e/2$ at a distance of its bond length (1.08~\AA), the Coulomb barrier of a neutral He atom located perpendicularly to the He$_2^+$ bond is reduced from 16.6 to 9.8~eV when the distance between the He and the He$_2^+$ decreases from 3.6 to 1.9~\AA. When the NIR laser electric field is additionally applied, the barrier of 9.8~eV is further reduced to 7.6, 6.4 and 5.2~eV at the pulse peak for intensities of $2\times10^{13}$, $5\times10^{13}$ and $10^{14}$~Wcm$^{-2}$, respectively. 
Since EII is the main ionization channel, for an estimate of the feasibility of ionization the barrier heights can be compared with the maximum kinetic energy a free electron acquires during a laser cycle (twice the ponderomotive energy). For the NIR photon energy of the experiment, 1.56 eV, the maximum electron kinetic energy is 2.4, 5.9 and 11.8~eV for intensities of $2\times10^{13}$, $5\times10^{13}$ and $10^{14}$~Wcm$^{-2}$, respectively, so that EII of He ligand atoms should be feasible at least for pulse intensities of $10^{14}$~Wcm$^{-2}$. Taking into account that in the presence of ions the laser field can accelerate the electrons to even higher energies~\cite{Saalmann:2003}, the pulse peak intensity requirements are expected to be even lower than $10^{14}$~Wcm$^{-2}$. Additionally, the positively charged snowballs likely retain at least part of the photoelectrons in bound or quasi-bound states in the droplets. These retained electrons act can as seed electrons for the EII avalanche and therefore further enhance the susceptibility of the droplet for laser-induced ignition. Thus, due to the reduced Coulomb barriers
one may expect that the presence of just a few snowball complexes and electrons in a droplet can keep the droplet susceptible for ignition by a NIR pulse long after XUV irradiation. 

Alternative scenarios could be the formation of long-lived He$^*$ and He$_2^*$ excitations in He droplets by recombination of photoelectrons and ions~\cite{buchenau1991excitation,mauracher2018cold}, as well as electrons bound to droplets in metastable bubble states~\cite{jiang1993electron,farnik1998differences}. 
Indeed, even at the initial photoelectron kinetic energies up to 8~eV the photoelectron can get captured in large He droplets by multiple electron-He scattering and subsequent bubble formation~\cite{henne1998electron}. However, the formation of metastable electron bubbles in He droplets in the presence of ions appears unlikely due to strong Coulomb attraction. Long-lived He$_{1,\,2}^*$ excitations, either in highly excited Rydberg states or in low-lying metastable states correlating with the $1s2s\,^{1,\,3}S$ atomic states, might be formed by electron-ion recombination in He nanodroplets. We mention that highly excited Rydberg states have been observed in pure and doped He nanodroplets with lifetimes in the nanosecond range~\cite{vonHaeften2005,loginov2011unusual}. However, highly excited He droplets are known to rapidly decay by autoionization~\cite{peterka2003photoelectron,asmussen2021unravelling} and by electronic relaxation leading to the emission of Rydberg atoms into the vacuum~\cite{kornilov2011femtosecond,mudrich2020ultrafast,asmussen2021unravelling}. He$_{1,\,2}^*$ excitations have not been observed at photon energies exceeding 26~eV in previous experiments using tunable XUV radiation, see the SM. Besides, even if He$_{1,\,2}^*$ excitations were formed in a He droplet, they would rapidly decay by interatomic Coulombic decay~\cite{ovcharenko2020autoionization,laforge2021ultrafast}. Thus, only the formation of stable snowball complexes appears to be a viable explanation for the experimentally observed long-lasting ignition signal, as detailed in the following. However, we cannot strictly rule out other processes.


In an attempt to simulate the dynamics occurring at very long pump-probe delays up to several nanoseconds, for which we still detect an enhanced ignition probability in the experiment, we use a modified version of the MD simulation that includes implanted rigid He$_{13}^+$ snowball complexes in the droplets. Since at present the simulation neglects interactions between neutral atoms and ions, it cannot describe the snowball formation process and its consequences on the evolution of the entire droplet such as evaporation of He atoms caused by the release of heat of snowball formation. Also, the small time steps of $4\times10^{-4}$~fs needed for an accurate propagation of the electron trajectories prohibit simulations extending to nanoseconds. Therefore we apply additional approximations and simplifications. 
In this second type of simulations, we initially implant a small number $N_\mathrm{snow}$ of icosahedral He$_{13}$ subunits in a He$_{10149}$ droplet, cutting out the original He atoms in the droplet such that a minimum distance of 3.6~\AA~between snowball atoms and original droplet atoms is preserved. Inside each He$_{13}$ subunit, the distance between the central neutral He atom and its surrounding 12 neutral He ligand atoms is 1.9~\AA. Only the central He atom is then ionized at the beginning of the trajectory. Thus, the droplets initially contain [He$_{13}$]$^+$-electron pairs instead of He$^+$-electron pairs. While in our snowballs a He$^+$ ion forms the core, in real snowballs the core consists of a He$_2^+$ or He$_3^+$ molecular cation with a shell of neutral He at a distance of about 1.9~\AA~\cite{knowles_structures_1996}. The Coulomb barrier at the He ligand atoms is slightly lower, 9.5~eV, than for the value of 9.8~eV given above for a He$_2^+$ ion.

Despite this difference, our simplified description of a He$_n^+$ snowball accounts at least qualitatively for the lowering of the Coulomb barriers at the surrounding He shell, which likely is an important feature to assist the ignition of the EII avalanche driven by the NIR pulse. The simulations are carried out for various numbers of snowballs $N_\mathrm{snow} = 1,\,3,\,7,\,13$ which could be realized in large droplets as those used in this experiment. After the central He atom in the snowball is photoionized, the system is propagated for 800~fs with all inner ionization channels being switched off, so that the released photoelectrons equilibrate; $\approx2/3$ of them leave the droplet and stay partly in the vicinity of the droplet. The remaining photoelectrons roam through the entire droplet volume, occasionally visiting a snowball complex without bias towards the snowball of their parent He$^+$ ion. After this equilibration period, the inner ionization channels are enabled and the NIR probe pulse reaches its intensity maximum at $\Delta t = 1.4$~ps. The total neglect of interactions between ions and neutral atoms requires that nuclear coordinates are frozen; otherwise the He$_{13}^+$ snowballs would decompose due to the Coulomb repulsion acting between He$^+$ ions of different snowballs long before the arrival of the NIR pulse. Consequently, Coulomb explosion cannot be accounted for by the simulations. 

In summary, in the `snowball version' of the simulation model, the formation process of snowball complexes inside the droplet is not simulated, but the He droplet is doped with a given number $N_\mathrm{snow}$ of snowball complexes in an \textit{ad hoc} manner. The preparation procedure of the doped droplet can be seen as an attempt to mimic the fully relaxed state of a He droplet reached $\gtrsim 100$~ps after its irradiation by an XUV pulse. Consequently, (i) the simulated pump-probe signal is time independent, because the irradiation with the NIR pulse is not associated with a specific instant of the droplet evolution. (ii) Since the number $N_\mathrm{snow}$ of snowball complexes is an input parameter of the simulation, the simulated ignition probabilities cannot be related to those of the ps pump-probe signals shown in Fig.~\ref{fig:pssi}. Such a direct relation would be possible only if the snowball formation were explicitly included in the simulation, so that $N_\mathrm{snow}$ followed naturally from the number $N_\mathrm{pump}$ of He$^+$-electron pairs initially generated by the XUV pump pulse. Helium ions formed in He nanodroplets by photoionizaton or electron impact are known to be ejected to some extent by a non-thermal process~\cite{callicoatt1998fragmentation}. However, as the droplets grow larger, the number of He ions that remain bound to the droplets increases. To our knowledge, the exact ratio of bound to ejected He ions as a function of droplet size is yet unknown. Thus, the current simulations do not allow for a smooth transition from the ps pump-probe signals shown in Fig.~\ref{fig:pssi} to the ignition probabilities obtained by the `snowball version' of the simulation. 
The snowball model only informs about the enhancement of the ignition probability of a droplet for a given size, number of snowballs and NIR-pulse intensity. A simulation capable of modelling the full dynamics including ultrafast pump-probe dynamics and formation of snowballs and their interaction with laser pulses is of great interest and subject of future work.


\begin{figure}
	\center
	\includegraphics[width=0.7\columnwidth]{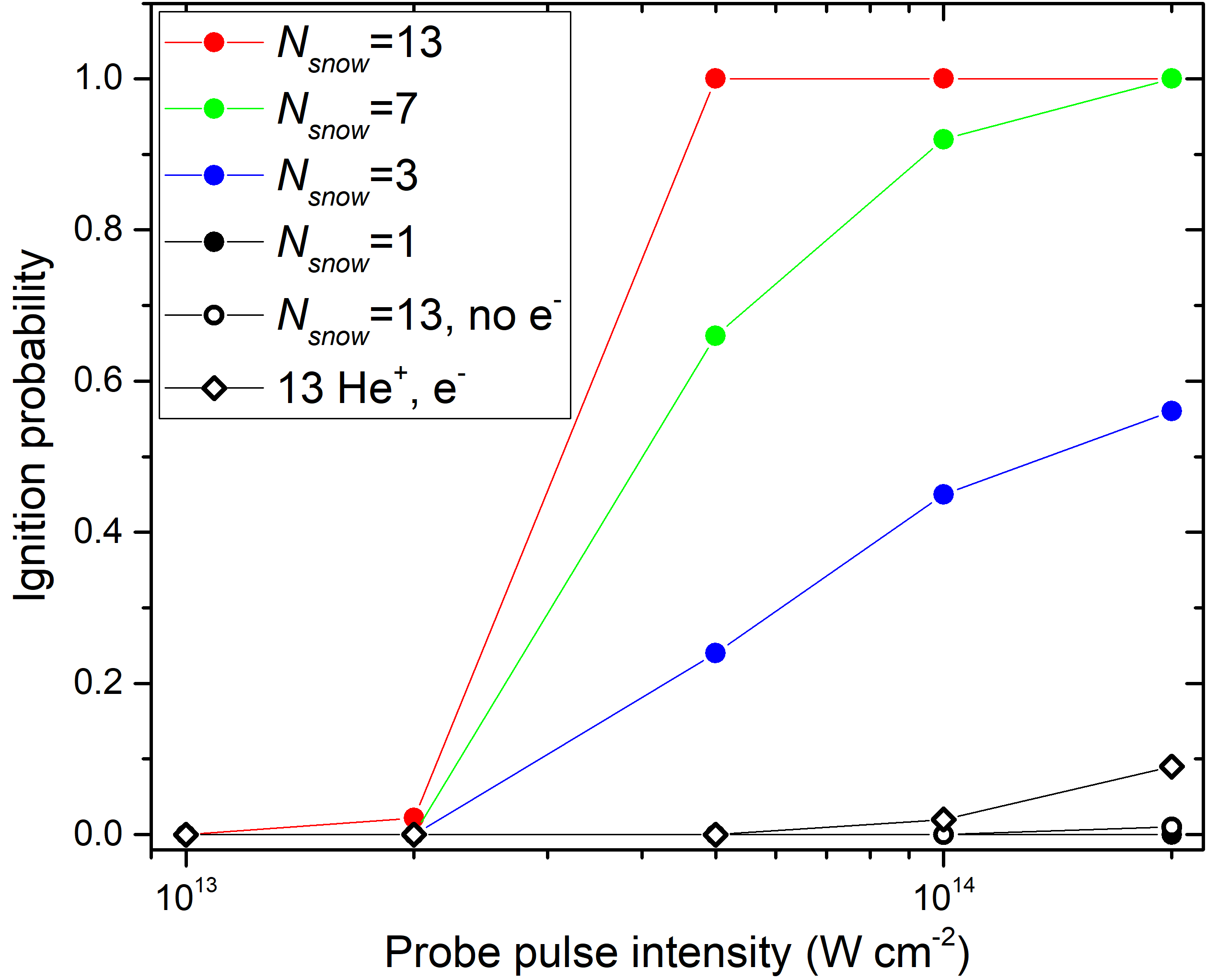}
	\caption{\label{fig:nssi} 
	Simulated ignition probabilities as a function of probe pulse intensity. Filled symbols: He droplets are initialized with a given number $N_\mathrm{snow}$ of pre-configured snowball complexes in each He droplet.
	Open circles: Droplets with 13 snowball complexes but no electrons. Open squares: Droplets with 13 He$^+$ ions and electrons with the He ions not being arranged as snowball complexes. See text for details. 
	}
\end{figure}


The ignition probabilities obtained in this version of the simulation are shown in Fig.~\ref{fig:nssi} as a function of the NIR probe-pulse intensity and for various numbers $N_\mathrm{snow}$ of snowballs implanted in the He nanodroplet (filled circles). Thus, it takes NIR intensities $>2\times 10^{13}$~Wcm$^{-2}$ to avalanche ionize a He nanodroplet containing $N_\mathrm{snow} > 1$ snowballs. This is in good agreement with the lowest NIR intensity at which we observe ignition of the He droplets in the experiment; At a minimum pulse energy of 60~$\mu$J (not shown), the NIR intensity averaged over the focal spot of the size one beam radius amounts to $3\times 10^{13}$~Wcm$^{-2}$. For $N_\mathrm{snow}\leq 1$, no ignition is predicted by the simulation, whereas in the experiment we do observe a low level of ignition at $t<0$ where no charges are present in the droplets. In the experiment, larger droplets were used , which feature a significant cumulative tunnel ionization probability (one of the He atoms tunnel ionizes when subjected to the NIR pulse). Additionally, large droplets tend to pick up molecules from the residual gas (mainly H$_2$O) owing to their large pick up cross section which further enhances their ignition probability. 

The simulation shows that EII by far ($>95$~\%) dominates the avalanche ionization channel, as it was also the case in all previous studies of pure and doped He droplets~\cite{heidenreich_charging_2017,schomas_ignition_2020,heidenreich_efficiency_2016}. The photoelectrons act as seed electrons for the EII avalanche. Accordingly, comparative simulations where the photoelectrons are discarded result in drastically reduced ignition probabilities. This is shown in Fig.~\ref{fig:nssi} by the open circles representing the results for $N_\mathrm{snow} = 13$. The ignition probability amounts to only 1~\% at most. Without photoelectrons, seed electrons are provided by TI which in our simulations occurred only in 1\,\% of trajectories at the highest considered intensity $I=2\times 10^{14}~\mathrm{Wcm}^{-2}$. 
This intensity is just below the threshold of $\sim5\times 10^{14}~\mathrm{Wcm}^{-2}$ at which TI triggers ignition in pristine droplets of size $N\sim10^4$ with unity probability.
Another set of comparative simulations (open squares in Fig.~\ref{fig:nssi}) is carried out for 13 photoelectrons and He$^+$ ions; However, these ions and electrons were not organized as snowball complexes with shortened interatomic distances to surrounding He ligand atoms (1.9~\AA), but at fixed interatomic distances as in the neutral droplet (3.6~\AA). (The same equilibration time was used and positions of all nuclei are kept fixed as in all other simulations of this section.) These simulations show higher ignition probabilities than for the case where snowballs are formed around the He$^+$ ions but photoelectrons are discarded (open circles). However, ignition probabilities are still much lower than for droplets containing both snowballs and electrons (closed circles). This indicates that the simultaneous presence of electrons and snowballs, the latter containing He atoms with lowered Coulomb barriers, are responsible for droplet ignition at nanosecond pump-probe delays.

With increasing numbers of snowballs $N_\mathrm{snow}$ an increasing number of electrons are retained in the droplet. At the end of the 800~fs equilibration phase (during which inner ionization is disabled), the electron population in the droplet amounts to 0.5 for $N_\mathrm{snow} = 3$ and 5 electrons for $N_\mathrm{snow} = 13$ on the trajectory-set average. Part of these electrons are trapped in the neighborhood of the snowballs; the trajectory-set averaged number of trapped electrons increases from 0.1 for $N_\mathrm{snow} = 3$ to 3 for $N_\mathrm{snow} = 13$. 
70-90~\% of the ionization avalanches induced by the NIR pulse can be traced back to snowballs; The vast majority of ionizations originate at snowballs which temporarily trapped electrons during the laser-free period. In the remaining cases, the ionization avalanche started somewhere else in the droplet or could not be unambiguously identified. 

When drawing conclusions from the simulation results, one must keep in mind the shortcomings of the simulation model. 
In particular, the formation of extended void bubbles that can trap electrons~\cite{henne1998electron} cannot be accounted for by our current classical model with fixed nuclei. What we infer from the present simulations with fair confidence is the basic principle that both electrons and He snowballs together induce the nanosecond pump-probe ignition signal, while the nature of the electron reservoir, \textit{e.~g.}, electrons trapped at snowballs or in bubbles, may be somewhat different in real large droplets. 


\section{Conclusions}
In conclusion, we have presented systematic XUV-pump and NIR-probe measurements of the rate of avalanche ionization of pure, relatively large He nanodroplets using high-harmonic pulses. Despite the low intensity of the XUV pulses compared to NIR pulses and soft x-ray pulses used to activate the droplets for avalanche ionization in earlier experiments, a high pump-probe signal contrast was achieved; Under certain conditions, ignition of the He nanodroplets by the NIR pulse was controlled by the XUV pulse with nearly unity contrast and a clear maximum of the ignition rate was observed at a pump-probe delay of 200~fs. The pump-probe dynamics at short delays on the ps scale is well reproduced by classical MD simulations. Mainly the expansion of the He photoions and the emission of XUV photoelectrons out of the He droplets determines the observed drop of the pump-probe ignition signal. 

The salient feature of these experiments is the persistent increased ignition probability for extremely long XUV-pump and NIR-probe delays up to nanoseconds. It is explained by the formation of stable ionic snowball complexes inside the He nanodroplets, which retain a number of electrons inside the droplets. Alternative scenarios where electrons trapped in bubbles or metastable or highly excited Rydberg states are attached to the droplets appear unlikely. Simplified MD simulations of the snowball system indicate that the combined action of these electrons and the snowballs, in which the ionization threshold of the neutral ligand atoms are lowered, crucially facilitate NIR-driven avalanche ionization. Future model calculations should include all relevant interactions between neutral and charged particles thereby encompassing the full dynamics from the formation of long-lived fragments and complexes up to nanoplasma ignition. The demonstrated ability of XUV irradiation to long-term activate nanodroplets for subsequent laser-driven avalanche ionization could potentially open up new routes to laser processing of nanostructures, soft-matter and surface systems.


\subsection{Acknowledgments}
The authors acknowledge ELI Beamlines, Doln\'{i} B\v{r}e\v{z}any, Czech republic, for the provided beamtime and thank the facility staff for their assistance and the Institute of Physics of the Czech Academy of Sciences for their support. This work was supported by the project ``Advanced research using high-intensity laser-produced photons and particles'' (ADONIS) (CZ.02.1.01/0.0/0.0/16 019/0000789) from the European Regional Development Fund and the Ministry of Education, Youth and Sports. Part of this research was funded by the Extreme Light Infrastructure ERIC. C. M., M. M. and F. S. gratefully acknowledge financial support by Deutsche Forschungsgemeinschaft (DFG) within the project MU 2347/12-1 and STI 125/22-2 in the frame of the Priority Programme 1840 ‘Quantum Dynamics in Tailored Intense Fields’ and by COST Action CA21101. M. M. is thankful for support by the Carlsberg Foundation. A. H. acknowledges financial support from the Basque Government (project IT1584-22); Computational and manpower support provided by IZO-SGI SG Iker of UPV/EHU and European funding (EDRF and ESF) is gratefully acknowledged. This publication is also based upon work of COST Action CA21101 ``Confined molecular systems: From a new generation of materials to the stars'' (COSY) supported by COST (European Cooperation in Science and Technology).

\section*{References}

\bibliographystyle{ieeetr}
\bibliography{BibCris}

\clearpage

\end{document}


\title{Supplementary Information\\
Long-lasting XUV-induced ignition of avalanche ionization of helium nanodroplets}

	\author{C. Medina \textit{et al.}}
	\date{\today}
	
	\maketitle

\renewcommand\refname{Supplementary References}
\renewcommand{\figurename}{Supplementary Fig.}

\begin{figure}[h!]
\centerline{\resizebox{0.6\textwidth}{!}{\includegraphics{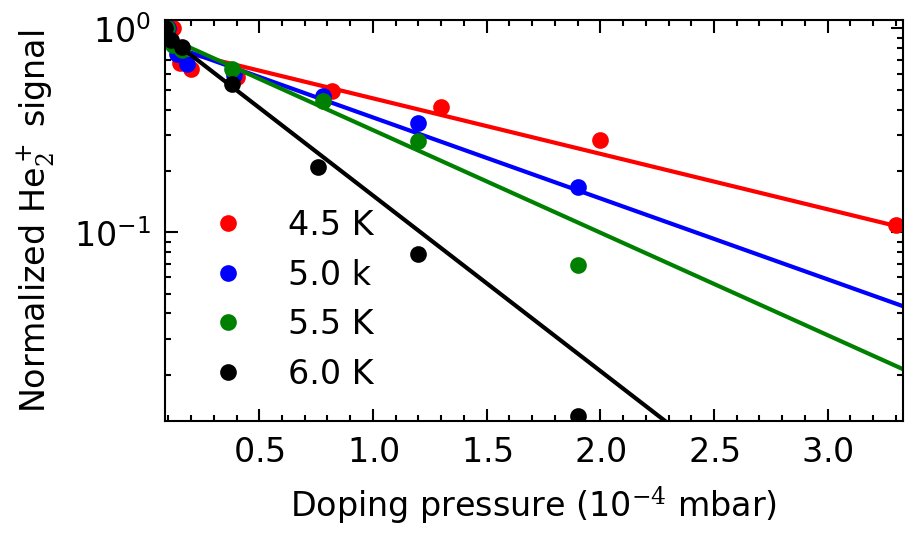}}}\caption{\label{fig:titration} Normalized He$_2^+$ signal measured in the last vacuum chamber of the He nanodroplet apparatus as a function of the He gas pressure in the doping chamber for different nominal temperatures of the He nozzle. Filled symbols represent the experimental values. Solid lines are linear fits in the single logarithmic plot, from which we determine the values of $\alpha$ according to Eq.~(\ref{eq:alpha}).}
\end{figure}

\section{Determination of the helium nanodroplet size}
The mean size of the He droplets $\langle N\rangle$ used in the experiment is determined by so-called titration measurements as described in Ref.~\cite{gomez_sizes_2011}. To this end, we leaked He gas into the doping chamber of length $L=0.296$~m located between the source chamber and the MAC chamber, see Fig.~1 in the main text. When the beam of He droplets passes through the He atmosphere the droplet beam is attenuated by multiple collisions of the He droplets and the He atoms in the background gas. By measuring the flux of He droplets passing through this chamber, through the MAC chamber, and all the way into the last chamber containing a quadrupole mass spectrometer (QMS) for different pressures of the He gas in the doping chamber, we determine $\langle N\rangle$ using the formula~\cite{gomez_sizes_2011},
\begin{equation}
    \langle N\rangle=C\pi\left(\frac{3}{4n_\mathrm{He}}\right)^2\left(\frac{L}{k_\mathrm{B}T}\sqrt{\frac{v_D^2+v_M^2}{v_D^2}}\cdot\frac{E_M}{E_V}\cdot\frac{1}{\alpha}\right)^3.
\end{equation}
Here, $C=3.3$, $n_\mathrm{He}=0.022~$\AA$^{-3}$, $T=293~$K, $v_D=236$~m/s is the droplet beam velocity, $v_M=1360$~m/s is the thermal velocity of the He atoms in the gas, $E_M=3k_\mathrm{B}T/2 + m_\mathrm{He}v_D^2/2 = 3765$~J/mol is the energy deposited in a He droplet upon collision with one He atom, and $E_V=73.3$~J/mol is the energy dissipated by evaporation of one He atom from a droplet. $\alpha$ is determined from the titration measurements shown in Supplementary Fig. 1 by
\begin{equation} \label{eq:alpha}
    \alpha\cdot P_M=\ln{\left(\frac{\Phi_\mathrm{He}(0)}{\Phi_\mathrm{He}(P_M)}\right)},
\end{equation}
where $P_M$ is the He pressure in the doping chamber and $\Phi_\mathrm{He}$ is the flux of He droplets measured by recording the He$_2^+$ signal using a quadrupole mass spectrometer QMS installed in the last chamber of the He droplet beam line. For the read-out temperatures of the He droplet nozzle of 4.5, 5.5, 5.5~K, we find $\alpha= 59$, $91$, $115$~Pa$^{-1}$m$^{-1}$ and thus $\langle N\rangle\approx 7\times 10^5$, $2\times 10^5$, $9\times 10^4$ atoms per droplet.

\section{Discussion of the nuclear and electron dynamics underlying the femtosecond-picosecond pump-probe signals}

\begin{figure}[h!]
\centering 
\centerline{\resizebox{0.5\textwidth}{!}{\includegraphics{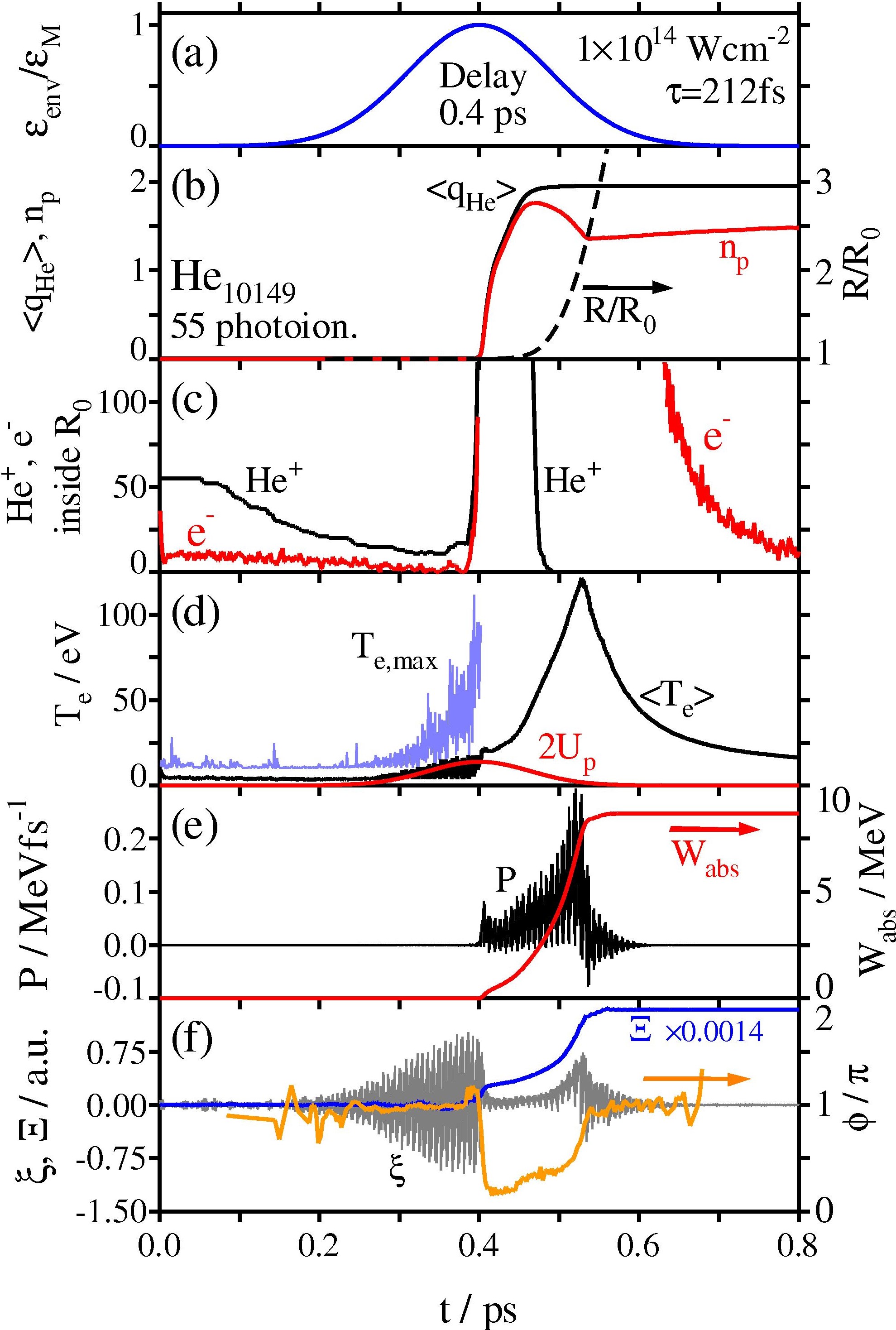}}}\caption{A single-trajectory example for the time evolution. (a) The normalized electric field envelope function of a Gaussian NIR pulse centered at $t = 0.4$~ps. The instant of the XUV photoionization is at $t = 0$. (b) The average bare He charge $\langle q_\mathrm{He}\rangle$ (without TBR, black closed curve) and the number $n_p$ of nanoplasma electrons per atom inside the nuclear framework of the expanding droplet (electrons within the distance of the most distant He ion from the center of mass of the droplet nuclear framework). The normalized droplet radius $R/R_0$ of the expanding nanoplasma (black dashed curve). Unlike for the determination of $n_p$, here $R$ and $R_0$ are taken from the median distances of He atoms or ions from the droplet center of mass. In this way, single photoionized He$^+$ ions kicked off prior to the ionization avalanche are excluded from the cluster expansion. (c) The number of electrons (red curve) and the number of He$^+$ ions within the unexpanded droplet of radius $43.1$~\AA~(red curve). (d) The average kinetic energy $\langle T_e\rangle$ per electron (black curve) and the maximum kinetic energy $T_{e,\textrm{max}}(t)$ of a single electron (light blue curve). $T_{e,\textrm{max}}(t)$ is displayed only until $t = 0.4$~ps, as subsequently $T_{e,\textrm{max}}(t)$ exceeds the range of values displayed in the panel. The maximum kinetic energy $2U_p$ (red curve) of a free electron (without cluster potential) is included for comparison, indicating that the quasi-free electrons of the nanoplasma are accelerated by a more efficient mechanism than by the laser electric field alone. (e) The power absorption $P$ (black curve) and energy absorption $W_\mathrm{abs}$ (red curve). (f) The reduced power absorption function $\zeta$ (Eq. (5), gray curve), the reduced energy absorption $\Xi$ (Eq. (6), blue curve) and phase shift $\phi$ between the NIR pulse electric field and the center-of-mass motion of the nanoplasma electron cloud (orange curve).\label{fig:SM_1}}
\end{figure}

In what follows we discuss the nuclear and electron dynamics of the short-time pump-probe signals on the femtosecond-picosecond timescale in greater detail. Supplementary Fig.~2 exhibits a number of time-dependent key quantities which characterize the time evolution of a single-trajectory example. Panel (a) shows the normalized Gaussian electric field envelope function of the NIR probe pulse centered at the pump-probe delay time of $\Delta t = 400$~fs. Here, $t = 0$ marks the XUV photoionization, taken as an instantaneous event. Panel (b) describes the inner ionization avalanche, outer ionization and Coulomb explosion of the droplet in terms of the average bare He charge $\langle q_\mathrm{He}\rangle$ (without ion-electron recombination), the number $n_p$ of nanoplasma electrons per atom inside the nuclear framework, and the relative radius $R/R_0$ of the expanding nuclear framework normalized to the initial droplet radius $R_0 = 43.1$~\AA. The NIR pulse almost completely inner ionizes the droplet (final average He charge of $\approx2$), which is always the case for these pulse parameters ($I_M = 10^{14}$~Wcm$^{-2}$, $\tau = 212$~fs, $\Delta t \geq 0$), once an inner ionization avalanche is triggered. EII is the dominating inner ionization channel (98\,\%). Only $\approx1/4$ of the electrons is stripped by outer ionization. The spatial expansion of the nanoplasma sets in notably at the peak of the NIR pulse.

Of particular interest is the initiation of the inner ionization avalanche, since this phase is decisive for the question whether the droplet contributes to the pump-probe signal. Since the photoelectrons are the seeds for the EII avalanche, panel (c) shows their number inside the unexpanded droplet of radius $R_0$. Included is also the number of He$^+$ ions. At $t=0$, photoionization generates 55 He$^+$-electron pairs. The number of electrons drops very rapidly to 20\,\% of its initial value within the first few femtoseconds, whereas the He$^+$ ions leave the droplets more slowly. With the onset of the inner ionization avalanche driven by the NIR pulse, the He$^+$ ions and electron population inside $R_0$ increases again and exceeds the definition range of the graph, to drop soon after in the course of the ionization to He$^{2+}$ and the Coulomb explosion.

To inquire the occurrence of nanoplasma resonances, in panels (d) – (f) the electron kinetic energy and laser energy absorption are monitored. In panel (d), $\langle T_e\rangle (t)$ is the average electron kinetic energy per electron, $T_{e,\textrm{max}}(t)$ the maximum kinetic energy of an electron at time $t$. Panel (e) shows the laser power absorption $P(t)$ and its time integral, the laser energy absorption $W_\mathrm{abs}$,
\begin{equation}
    W_\mathrm{abs}(t)=\int_{-\infty}^tP(t')dt',\,\,\mathrm{with}\,\,\sum_i eq_i\Vec{v}_i(t)\cdot\Vec{\varepsilon}(t).
\end{equation}
Here $q_i$ and $\Vec{v}_i$ are the charge in units of the elementary charge $e$ and the velocity vector of particle $i$, respectively, and $\Vec{\varepsilon}$ is the laser electric field. $P(t)$ is almost exclusively given by the contribution of the nanoplasma electrons. During the avalanche ionization, when $\langle q_\textrm{He}\rangle$ and $n_p$ rise steeply (panel (b)), $P$ and $W_\mathrm{abs}$ exhibit the typical resonance behavior which was observed in previous nanoplasma studies~\cite{Peltz.2011,Krishnan_2012,heidenreich_charging_2017}: Pronounced oscillations of $P$ and a steep rise of $W_\mathrm{abs}$, the latter converging shortly after the completion of inner ionization, and the average electron kinetic energy $\langle T_e\rangle$ reaching its maximum when $W_\mathrm{abs}$ converges to its final value. During this nanoplasma resonance, the maximum single-electron kinetic energy $T_{e,\textrm{max}}$ (light blue curve in panel (d)) assumes several hundred eV and exceeds the definition range of the panel. As a further indicator of resonance, we determine the phase shift $\phi(t)$ between the oscillating NIR laser electric field and the center-of-mass motion of the nanoplasma electron cloud~\cite{Saalmann_Ionization2003}. $\phi(t)$ is displayed as the orange curve in panel (f). Near the beginning and at the end of the ionization avalanche, $\phi(t)$ passes the value $\pi/2$, and assumes values not far from $\pi/2$ during the avalanche, in accordance with the criterion of a resonance situation~\cite{Saalmann_Ionization2003,heidenreich_charging_2017}.

While the aforementioned resonance \textit{during} the ionization avalanche is a common phenomenon, the more interesting question is whether there is an additional resonance prior to the ionization avalanche, when the droplet contains only the seed electrons generated by the photoionization. Such a resonance could increase the kinetic energy of the seed electrons and could therefore play an important role in the initiation of the avalanche. An early small first resonance was observed in simulations of He droplets doped with small Xe clusters followed by the main resonance during the avalanche ~\cite{heidenreich_charging_2017}.

Indeed, the maximum electron kinetic energy $T_{e,\textrm{max}}(t)$ (light blue curve in panel d) increases considerably from $\approx10$~eV before the arrival of the NIR pulse to values of 50-100~eV in the time interval $0.28~\mathrm{ps}\leq t \leq 0.38$~ps. $T_{e,\textrm{max}}(t)$ also exceeds by far the maximum kinetic energy of $2U_p$ (red curve in panel (d)) a free electron (\textit{i.~e.}, outside a potential well) can have during a laser cycle. $U_p$ is the ponderomotive energy
\begin{equation}
    U_p(t)=\frac{e^2\varepsilon^2_{env}(t)}{4m_e\omega^2},
\end{equation}
with the electric field strength $\varepsilon_{env}(t)$ of the Gaussian laser envelope function, the electron mass $m_e$ and the laser frequency $\omega$. Despite the $T_{e,\textrm{max}}$ values being enhanced for $0.28~\mathrm{ps}\leq t \leq 0.38$~ps, neither $P(t)$ nor $W_\mathrm{abs}$ (panel (e)) show any features in the corresponding time interval. However, the power absorption is essentially determined by the scalar product $\Vec{v}_i(t)\cdot\Vec{\varepsilon}(t)$ of the electron velocities and laser electric field in Eq.~(1); in case of resonance the scalar products are accumulated periodically over time. Since $P(t)$ also depends on the number $n_e(t)$ of nanoplasma electrons, which strongly increases over time, and the strength of the laser electric field, we introduced a ``reduced power absorption function'' $\zeta (t)$ in our previous work~\cite{heidenreich_charging_2017},
\begin{equation}
    \zeta (t)\equiv\frac{P(t)}{n_e(t)\left|\Vec{\varepsilon}(t)\right|},
\end{equation}
where the power absorption $P(t)$ is normalized to the number of nanoplasma electrons $n_e(t)$ and to the absolute value of the instantaneous laser electric field $\Vec{\varepsilon}(t)$. In analogy to $W_\mathrm{abs}$, we also introduced the reduced energy absorption function
\begin{equation}
    \Xi(t)\equiv\int_{-\infty}^t\zeta(t')dt'.
\end{equation}
In this way, the effect of the scalar products $\Vec{v}_i(t)\cdot\Vec{\varepsilon}(t)$ on a possible nanoplasma resonance is better brought out also in cases when only a few electrons are involved as well as when the laser electric field is still weak at the rising edge of a pulse.

$\zeta(t)$ and $\Xi(t)$ are portrayed in panel (f). $\zeta(t)$ shows strong oscillations in the time interval $0.28~\mathrm{ps}\leq t \leq 0.38$~ps, but $\Xi(t)$ assumes values near zero, indicating nearly no net reduced absorption. This together with a phase $\phi(t)$ near zero let us conclude that an early weak resonance is absent. Instead, we attribute the high maximum electron kinetic energies to an ion-assisted acceleration of single electrons by the laser electric field (``powered flyby'')~\cite{saalmann_2008,Heidenreich_2017}. In this acceleration mechanism, an electron gains velocity in the potential well of an ion during a close encounter, so that during a short path of the electron the scalar product $\Vec{v}_i(t)\cdot\Vec{\varepsilon}(t)$ of an electron can assume high values, if the laser electric field has the right phase. In this way, the time integral of $P(t)$ can be intermittently high for short traces of an electron path, leading to high electron accelerations and decelerations, but with much less net energy absorption than in a regular resonance where $\Vec{v}_i(t)\cdot\Vec{\varepsilon}(t)$ assumes high values periodically. Thus, as stated above, one may view resonance as a special case of powered flyby, where the $\Vec{v}_i(t)\cdot\Vec{\varepsilon}(t)$ are accumulated.

\begin{figure}[h!]
\centerline{\resizebox{0.5\textwidth}{!}{\includegraphics{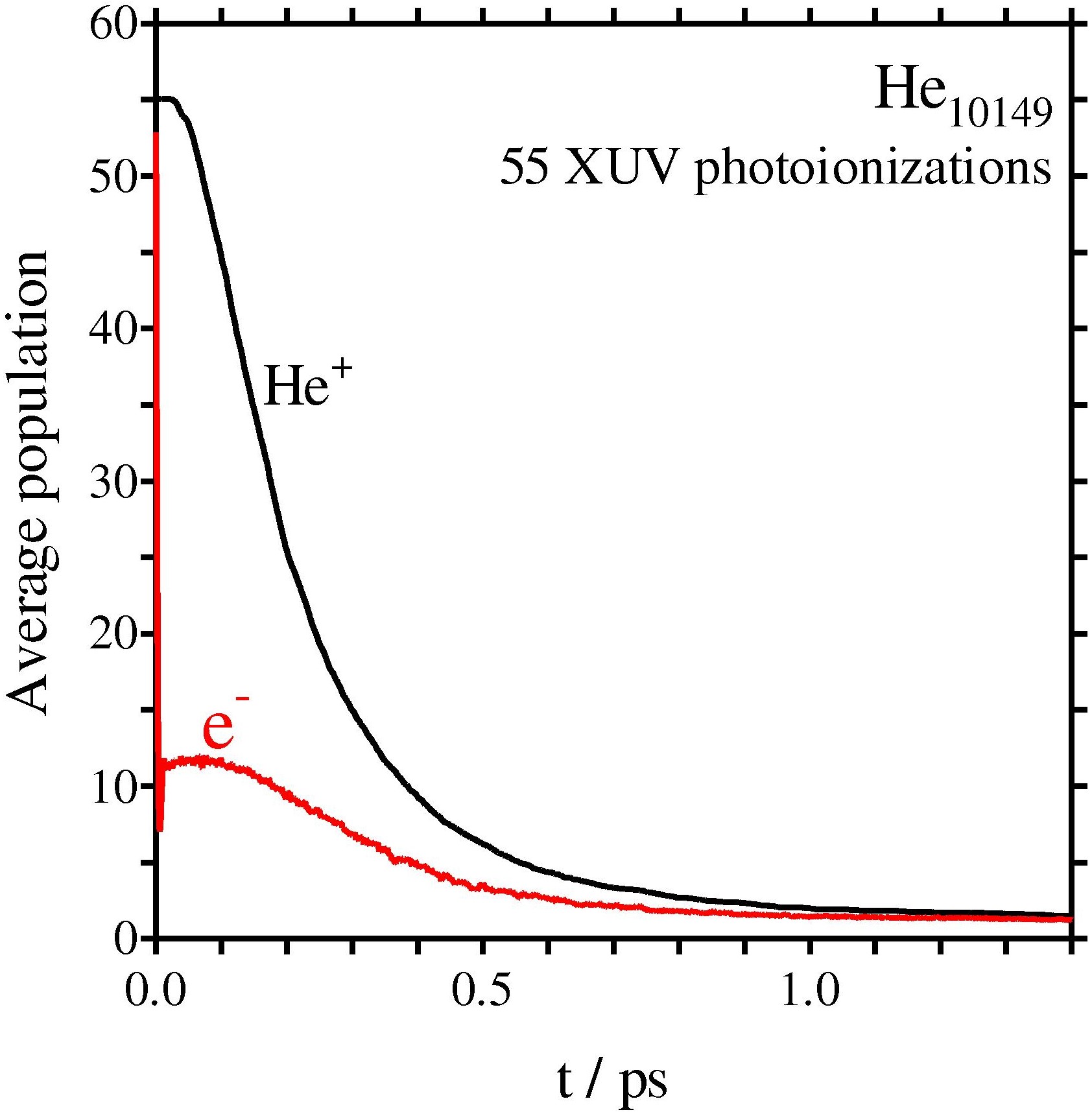}}}\caption{
The number of electrons (red curve) and the number of He$^+$ ions (black curve) inside the unexpanded He droplet of radius 43.1~\AA~following an instantaneous XUV photoionization of 55 He atoms. The number of electrons and He$^+$ ions is averaged over 100 trajectories.
\label{fig:SM_2}}
\end{figure}
Both powered flyby and resonance require the presence of cations in the droplet. While in Supplementary Fig.~2b the electron and He$^+$ populations were discussed in combination with a NIR probe pulse, Supplementary Fig.~3 shows the decay of the populations over the entire considered pump-probe delay time range of 1.4~ps without that a NIR pulse interferes. The populations are averaged over a set of 100 trajectories. While the time dependence of the He$^+$ population resembles a bell shape, the decay of the electron population consists of two components: (i) An almost sudden component of only several fs comprising $\approx80\,$\% of the photoelectrons, as in the single-trajectory example of Supplementary Fig.~2b. (ii) A slow bell-shaped component whose pace is similar to the one of the He$^+$ ions, obviously because the potential retaining the electrons flattens out with the escape of the He$^+$ ions from the droplet. At $t = 1.4$~ps, the droplet contains on average 1.5~He$^+$ ions and 1.3 electrons.

\begin{figure}[h!]
\centerline{\resizebox{0.5\textwidth}{!}{\includegraphics{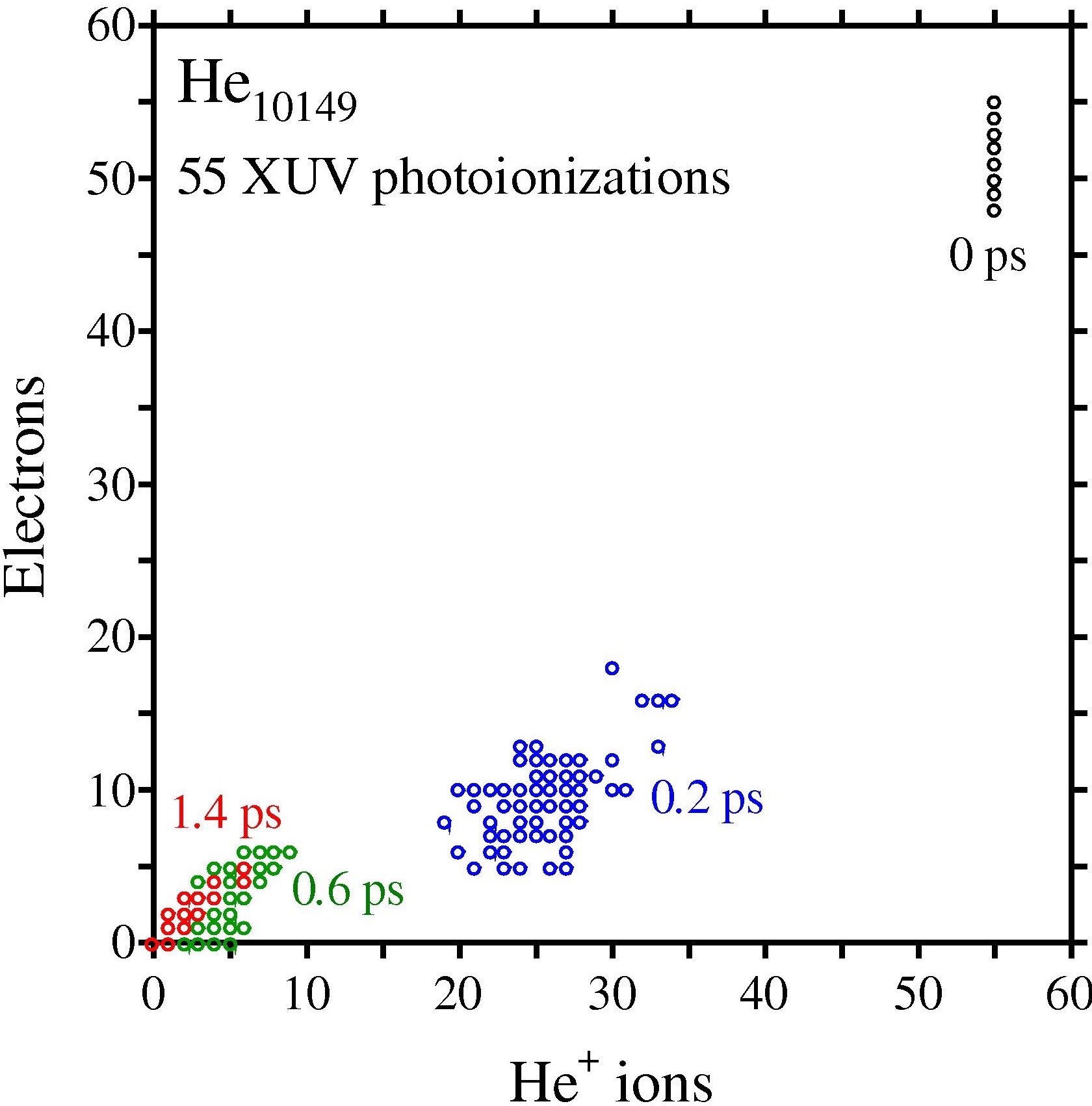}}}\caption{
Correlation plot of the number of electrons \textit{vs.} the number of He$^+$ ions inside the unexpanded He droplet of radius 43.1~\AA~following an instantaneous XUV photoionization of 55 He atoms. Data sets (100 trajectories per set) are given for times $t = 0,~0.2,~0.6$ and $1.4$~ps after photoionization, each data point representing a single trajectory. 
\label{fig:SM_3}}
\end{figure}

By how much these populations vary among the trajectory set is shown by the correlation plot in Supplementary Fig.~4 for times $t = 0,~0.2,~0.6$ and $1.4$~ps after 55 photoionizations at $t = 0$. For every trajectory, the number of electrons inside the droplet is plotted vs. the number of He$^+$ ions, every data point representing a trajectory. Due to the implementation of the photoionization as an instantaneous event, the droplet loses a few electrons even at $t = 0$, when an atom near the droplet surface is ionized and the initial random position of the photoelectron happens to be outside $R_0$. At $t = 1.4$~ps, the population varies between no ions and electrons and at most six He$^+$ ions and five electrons.

\begin{figure}[h!]
\centerline{\resizebox{0.5\textwidth}{!}{\includegraphics{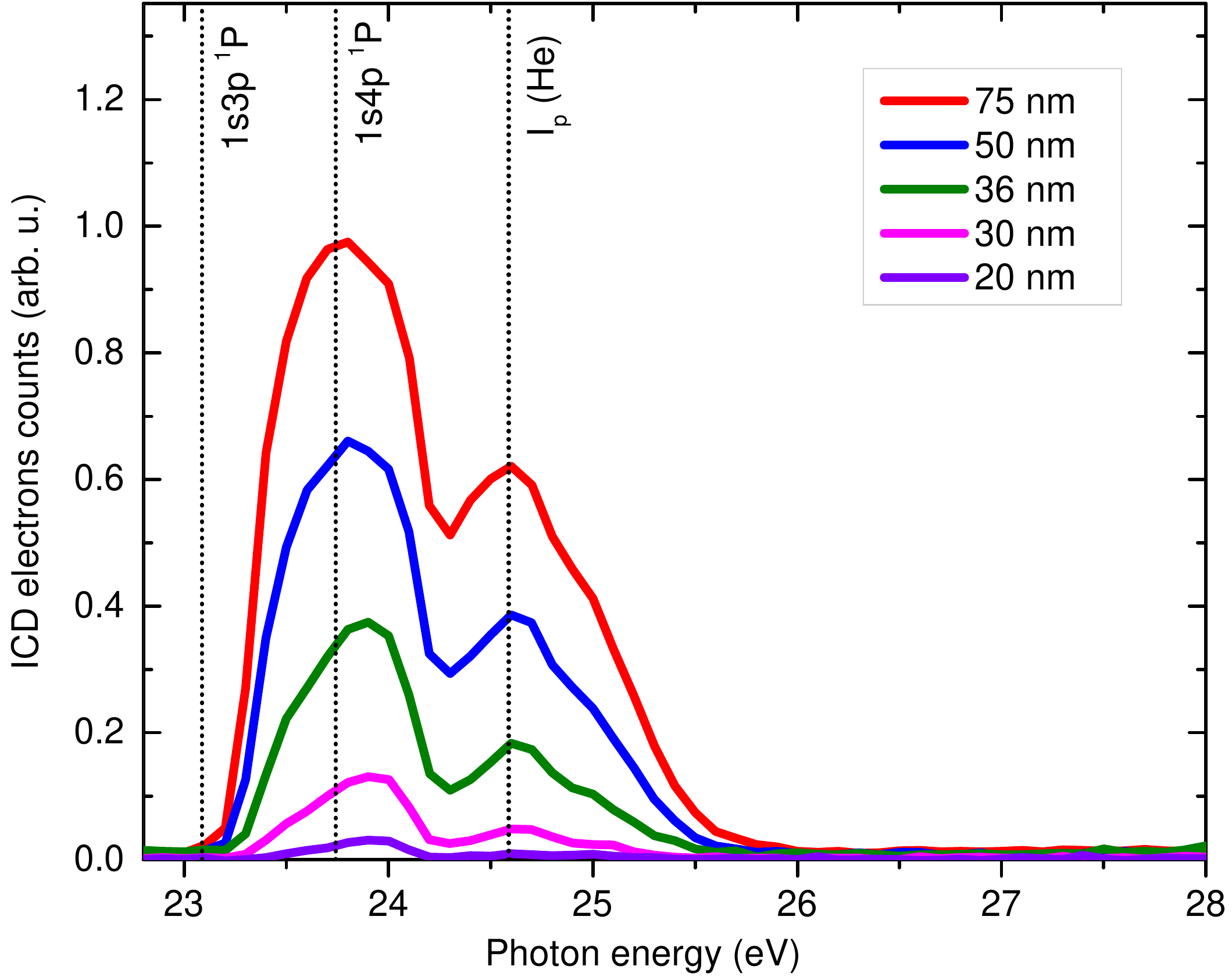}}}\caption{
Measurements of the yield of electrons created by Interatomic Coulombic Decay (ICD) of multiply excited He nanodroplets of given radius by tunable XUV radiation generated by the synchrotron radiation source Elettra, Trieste. 
\label{fig:SM_4}}
\end{figure}

\section{Resonant excitation of H{e} droplets}
To assess the relevance of long-lived excited states in He droplets in the present experiments we studied the decay of large He droplets that were resonantly excited using tunable XUV synchrotron radiation. Multiple excitations in one He droplet rapidly decay by the process $\mathrm{He}^* + \mathrm{He}^* \rightarrow \mathrm{He} + \mathrm{He}^+ + e_\mathrm{ICD}$ often called Interatomic Coulombic Decay (ICD)~\cite{Kuleff:2010,ovcharenko2020autoionization,laforge2021ultrafast}. The measured yield of characteristic ICD electrons $e_\mathrm{ICD}$ as a function of photon energy $h\nu$ is shown in Supplementary Fig.~5 for various radii of the He droplets. These spectra resemble those measured by detecting fluorescence emission for large He droplets~\cite{von2011size}. While high yields of ICD electrons are measured in the range 23-26~eV, no ICD of He$^*$ is found at $h\nu >26$~eV. We conclude that while electrons may be captured in large He droplets even at $h\nu>26$~eV by electron-He scattering and bubble formation, either short-lived He$^*$ states are formed which decay faster than ICD by other processes (autoionization~\cite{peterka_photoionization_2007}, desorption~\cite{kornilov2011femtosecond}), or more stable structures form such as snowballs and quasi-bound snowball-electron systems.

\section*{Supplementary references}

\bibliographystyle{ieeetr}
\bibliography{BibCris}